    \documentclass[preprint2]{aastex}

\usepackage[svgnames]{xcolor}
\usepackage[colorlinks=true,linkcolor=blue, citecolor=blue]{hyperref}

\shorttitle{Three new contact binaries?}
\shortauthors{Audrey Thirouin, and Scott. S. Sheppard}

\begin{document}

\title{The Plutino population: An Abundance of contact binaries.}

\author{Audrey Thirouin\altaffilmark{1}}
\affil{Lowell Observatory, 1400 W Mars Hill Rd, Flagstaff, Arizona, 86001, USA. }
\email{thirouin@lowell.edu}

\and
 
\author{Scott S. Sheppard\altaffilmark{2}}
\affil{Department of Terrestrial Magnetism (DTM), Carnegie Institution for Science, 5241 Broad Branch Rd. NW, Washington, District of Columbia, 20015, USA.}
 \email{ssheppard@carnegiescience.edu}

\begin{abstract} 

We observed twelve Plutinos over two separated years with the 4.3m Lowell's Discovery Channel Telescope. Here, we present the first lightcurve data for those objects. Three of them (2014~JL$_{80}$, 2014~JO$_{80}$, 2014~JQ$_{80}$) display a large lightcurve amplitude explainable by a single elongated object, but are most likely caused by a contact binary system due to their lightcurves morphology. These potential contact binaries have rotational periods from 6.3~h to 34.9~h and peak-to-peak lightcurve variability between 0.6 and 0.8~mag. We present partial lightcurves allowing us to constrain the lightcurve amplitude and the rotational period of another nine Plutinos. By merging our data with the literature, we estimate that up to $\sim$40$\%$ of the Plutinos could be contact binaries. Interestingly, we found that all the suspected contact binaries in the 3:2 resonance are small with absolute magnitude H$>$6~mag. Based on our sample and the literature, up to $\sim$50$\%$ of the small Plutinos are potential contact binaries.

\end{abstract}

\keywords{Kuiper Belt Objects: 1995~HM$_{5}$, 2001~KB$_{77}$, 2006~UZ$_{184}$, 2014~JK$_{80}$, 2014~JL$_{80}$, 2014~JO$_{80}$, 2014~JP$_{80}$, 2014~JQ$_{80}$, 2014~JT$_{80}$, 2014~KC$_{102}$, 2014~KX$_{101}$, 2015~BA$_{519}$, Techniques: photometric}

\section{Introduction}

Any object confined in a mean motion resonance with Neptune receive the name of \textit{resonant object}. At $\sim$39.4~AU, the 2:3 resonance is the most stable and densely populated resonance \citep{Chiang2002, Jewitt1998}. As Pluto is a 3:2 resonant object, all bodies trapped into this resonance receive the denomination of \textit{Plutinos}. According to \citet{Malhotra1995}, some Plutinos cross Neptune's orbit, but they never suffer a close approach. The overabundance of Plutinos is likely due to Neptune's migration \citep{Malhotra1993, Malhotra1995}. Neptune could have been created in the Inner Solar System and migrated outwards to its actual location, due to angular momentum exchange with surrounding planetesimals \citep{Fernandez1984}. With Neptune's migration, the resonances moved into the trans-Neptunian belt and the planetesimals were captured by such resonances \citep{Levison2008, Malhotra1995}. In conclusion, the migration and circularization of this planet's orbit provoked the capture of objects into the resonances. A migration of $\sim$8~AU over 10$^{7}$~years reproduces the observed distribution of Plutinos \citep{Gomes2000}.  

Several of the largest Plutinos are binaries or multiple systems: Pluto-Charon (and 4 smaller moons), Huya (formerly 2000~EB$_{173}$), Orcus-Vanth (2004~DW), Lempo (1999~TC$_{36}$, triple system), and 2003~AZ$_{84}$. These binaries are likely formed by collisional impact on the primary. Trapped in the Plutinos, Mors-Somnus (2007~TY$_{430}$) is a noteworthy equal-sized wide system that cannot have been created by impact, and is like a dynamically Cold Classical equal-sized binary \citep{Sheppard2012b}. Finally, 2001~QG$_{298}$ is a near equal-sized contact binary \citep{Sheppard2004}. Therefore, less than 3$\%$ of the Plutinos are known binaries. The Plutino population seems to have a deficit in separated wide binary systems compared to the other dynamical groups \citep{Noll2008, Compere2013}. As said in \citet{Thirouin2017}, contact binaries are unresolved with the \textit{Hubble Space Telescope (HST)}. These compact systems may remain undiscovered and therefore the current fraction of close/compact binary systems in the Plutino population and other populations by extension are unknown. Similarly, as argued by \citet{Porter2012}, there is potentially a large fraction of binaries  with circular and very tight orbit due to Kozai effects that are undetectable with \textit{HST}. Even if it is still unclear how contact binaries are formed, one can argue that tidal effects as suggested by \citet{Porter2012} can create very tight orbits, but the gravitational collapse is also an option \citep{Nesvorny2010}. Finally, as suggested by the star formation theory, if a triple system looses one component, the other two objects have to shrink their orbit to go back to a stable configuration and can potentially create a contact/close binary \citep{Bodenheimer2011}. However, it is important to point out that several of theses models are not dedicated to the TNO science, and that we are still missing several observables to infer if a model is working or not. We aim that our work regarding contact binaries will help to constrain their localizations, characteristics, fraction among other properties that can be used for modeling or as observables to check the validity of a model.  



 



\section{Data reduction, analysis, and results}
\label{sec:lightcurve} 

Data obtained between May and July 2017, and November-December 2015 with the Lowell Observatory's 4.3~m Discovery Channel Telescope (DCT) are presented. We used exposure times of 300 to 500 seconds with a broad-band filter with the Large Monolithic Imager (LMI) \citep{Levine2012}. Details about the telescope and instrument can be found in \citet{Thirouin2017}. The observing log is in Table~\ref{Tab:Log_Obs}. 

The data reduction and analysis is the same as in our previous works, but following we summarize the main steps. Images were biased subtracted and flatfielded thanks to calibration images obtained each night (dome and/or twilight flats depending on the weather conditions). Then, we proceed with the aperture photometry with a data reduction software using the Daophot routines described in \citet{Stetson1987}. The optimal aperture radius in order to estimate the flux of the object and limit the background contamination has been estimated by a growth curve \citep{Stetson1990, Howell1989}. We selected 20 stars per field as reference stars. We also played with the pointings in order to keep about the same field of view every night and thus the same reference stars or at least keep a couple of reference stars in common between pointings. Such a technique allows us to use the same reference stars across observing runs and also allow us to merge our datasets. The error associated to the photometry has been estimated following the formalism in \citet{Howell1989}. Finally, we looked for periodicities using the Lomb periodogram technique \citep{Lomb1976, Press1992}, and double-checked our results with the Phase Dispersion Minimization from \citet{Stellingwerf1978}. More details about our data reduction/analysis procedures are available in \citet{Thirouin2010, Thirouin2012, Thirouin2013}. 

In this section, we show our photometric results for our partial and complete lightcurves. All our lightcurves with a rotational period estimate are plotted over two cycles. In case of partial lightcurve without rotational period estimate, the photometry with error bars is plotted versus Julian Date (no light-time correction applied). We summarize our photometry in Appendix A, and our results are reported in Table~\ref{Summary_photo}.   

It is important to point out that our study was designed to constrain the percentage of large amplitude objects in the Plutino population. Therefore, if an object was not showing obvious sign of large variability characteristic of a contact binary, we stop observing it. But, in some cases our observations were interrupted due to bad weather and/or smoke. In conclusion, for the rest of our study, all objects with an observed variability smaller than 0.2~mag will be considered as a non contact binary.

\subsection{Lightcurves of potential contact binaries}

Now, the lightcurves of 2014~JL$_{80}$, 2014~JO$_{80}$, and 2014~JQ$_{80}$ are displayed. All of these objects present a large variability between 0.55 and 0.76~mag, and rotational periods from 6.32~h up to almost 35~h. These objects will be analyzed in detail in Section~\ref{sec:analysis}.  

 \subsubsection{2014~JL$_{80}$}

With H=7.1~mag as absolute magnitude according to the Minor Planet Center (MPC), we estimate a size of 253/113~km (geometric albedo of 0.04/0.20).

\begin{figure*}
\includegraphics[width=9cm, angle=0]{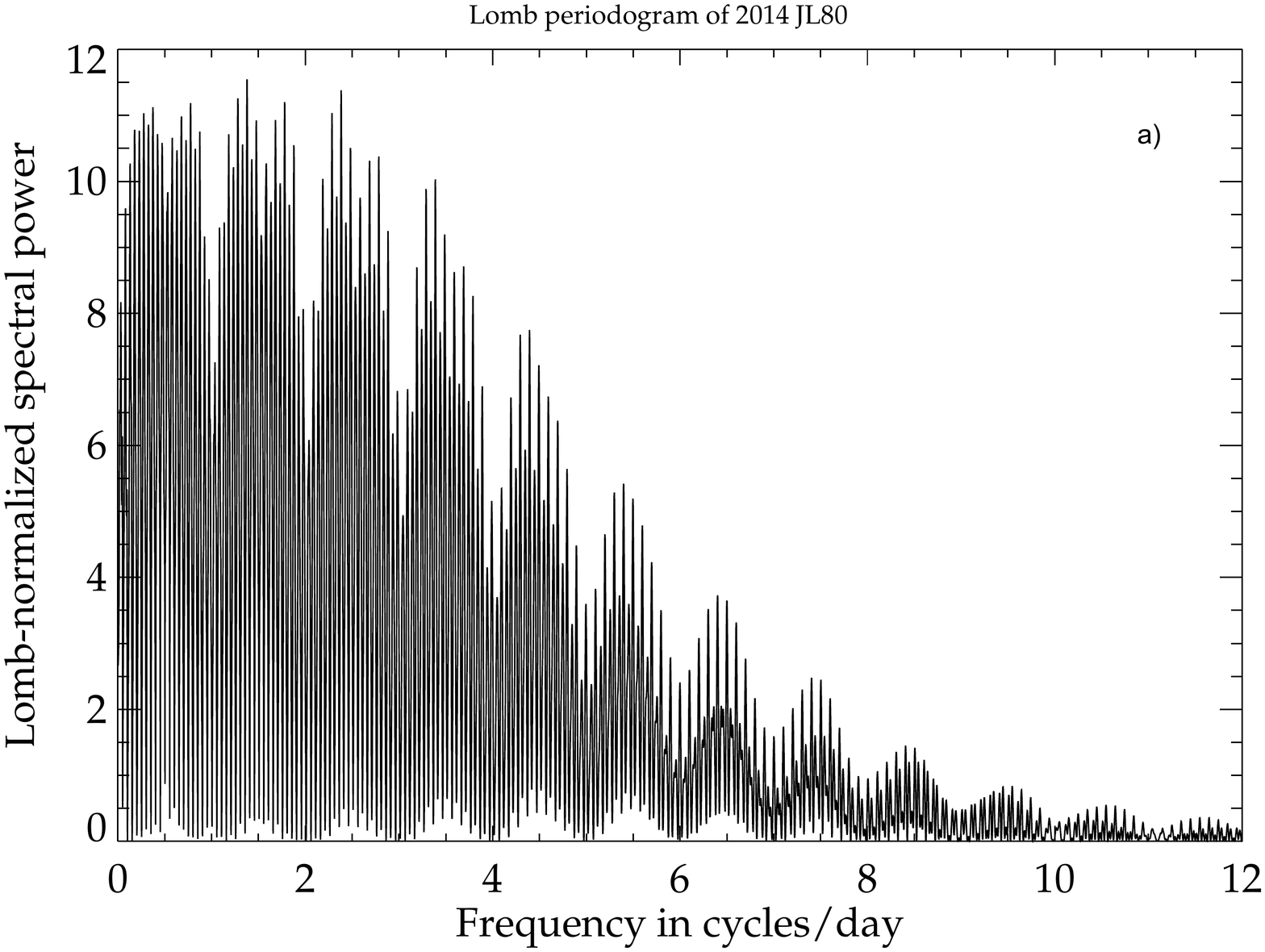}
\includegraphics[width=10cm, angle=0]{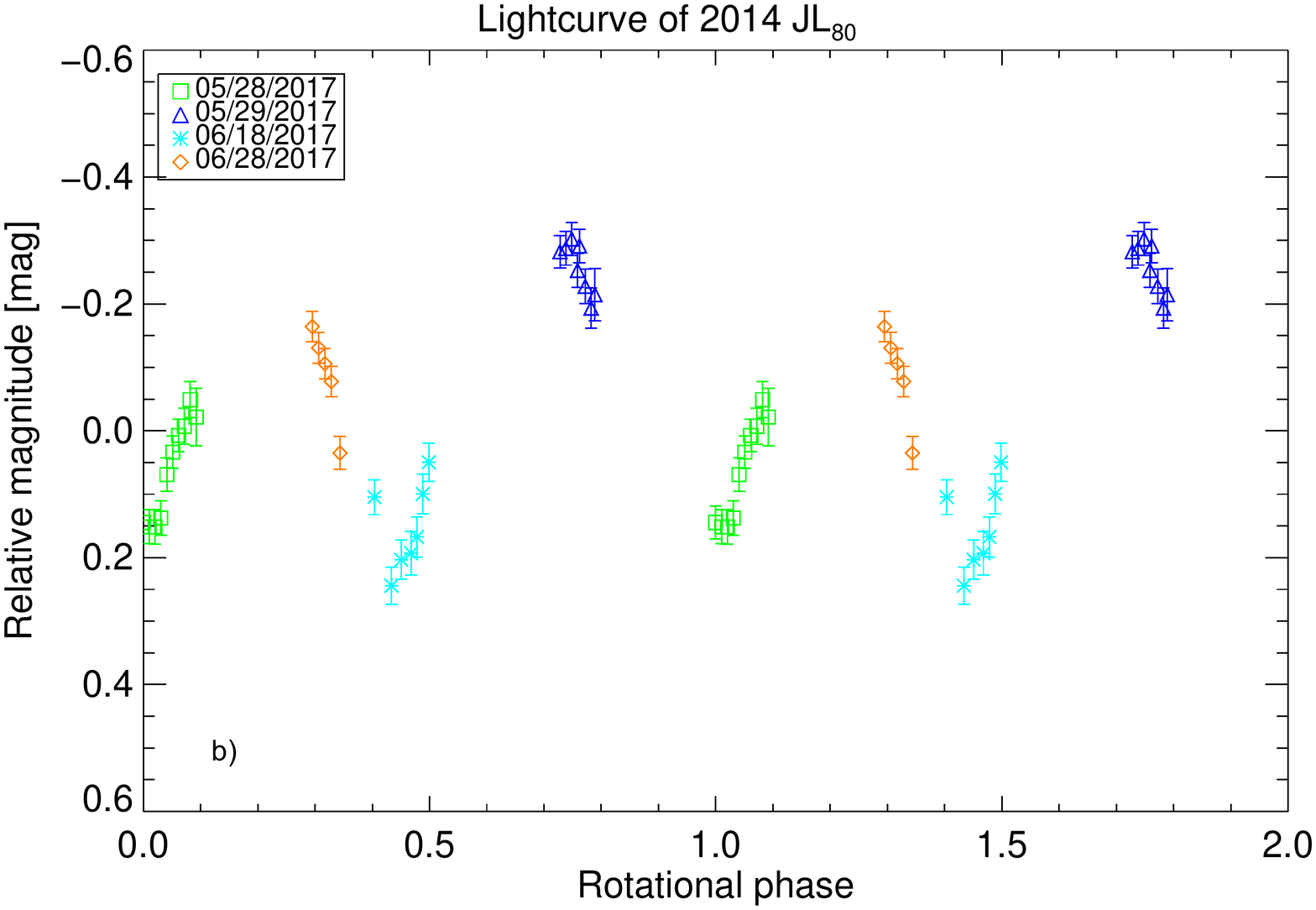} 
\includegraphics[width=9cm, angle=0]{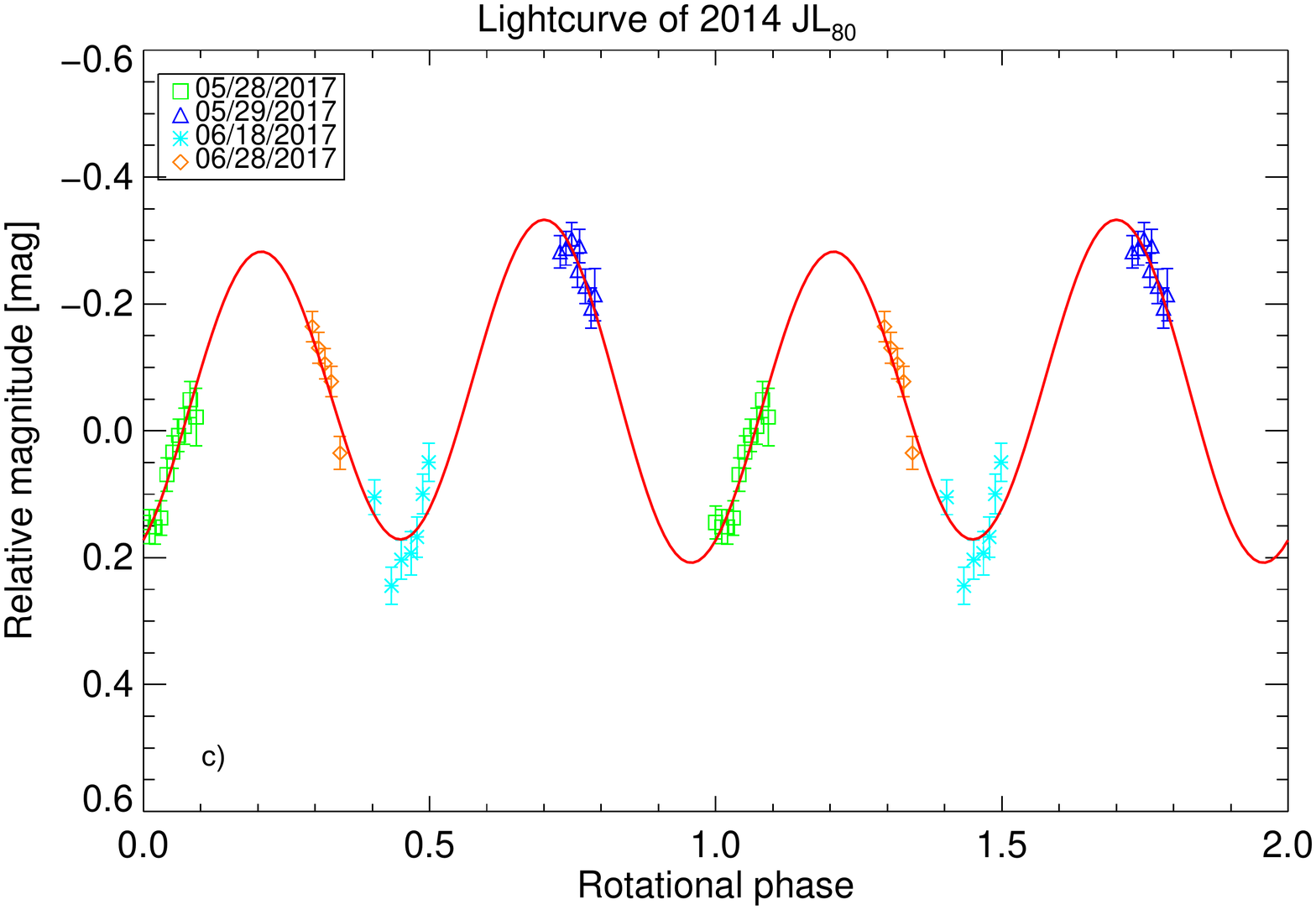}
\includegraphics[width=9cm, angle=0]{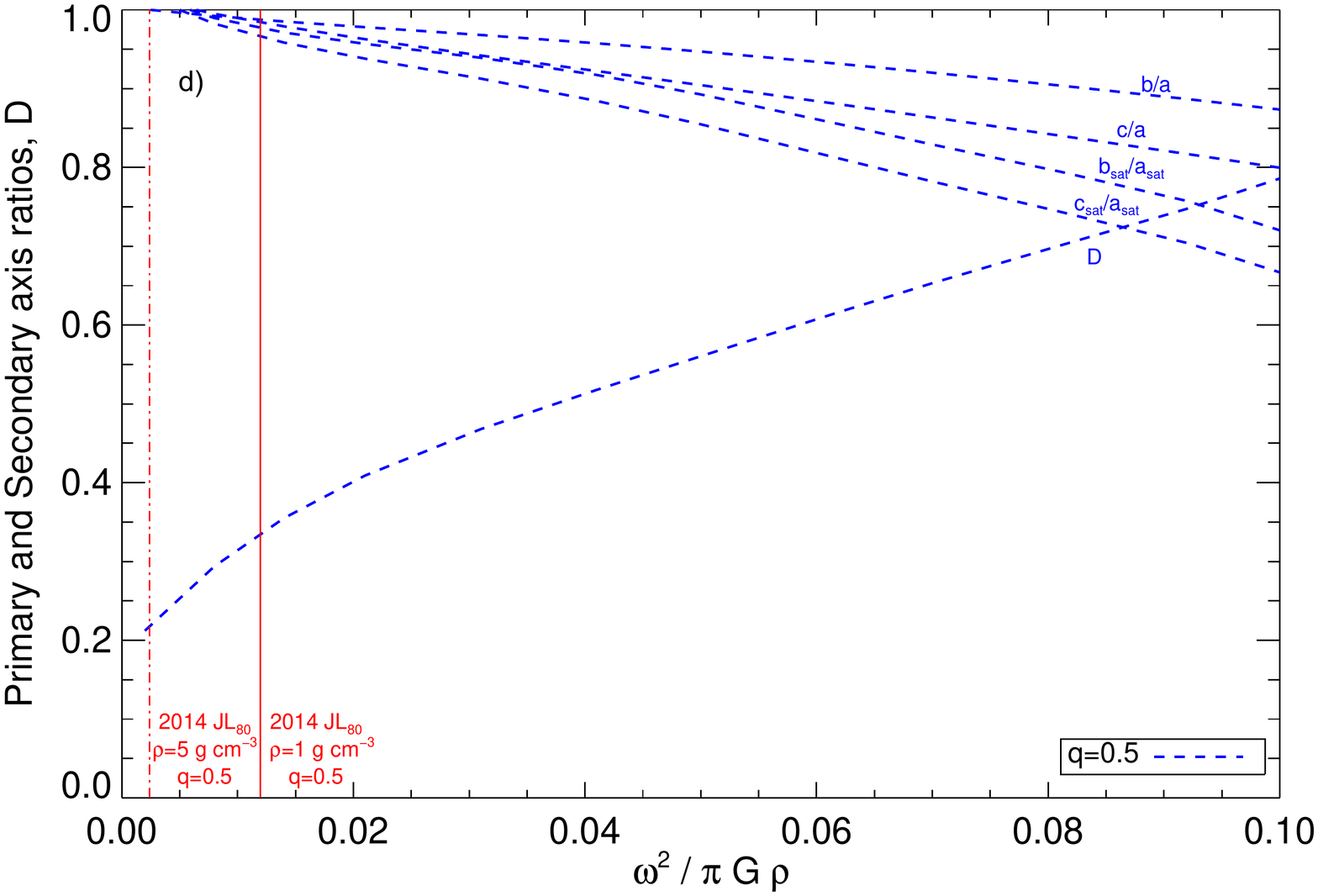}
\caption{\textit{Study of 2014~JL$_{80}$:} The Lomb periodogram presents several peaks, but the highest one suggests a double-peaked periodicity of 34.87~h (plot a)). The corresponding lightcurve is plotted on plot b). We tried to fit a Fourier Series (2nd order) to our data (red continuous line, plot c)), but the fit failed to reproduce the V-shape of the second minima. Because the lightcurve displays a large variability, and because of the V-shape of the second peak, we suggest that this object is a contact binary. The plot d) was used to derive the basic information of the system, assuming a contact binary nature.  }
\label{fig:JL80}
\end{figure*}

After two nights of observations of 2014~JL$_{80}$, a very long rotational period of more than 24~h, and a large variability were suspected. Three more nights with sparse sampling observations were required to present a potential lightcurve. The highest peak of the Lomb periodogram corresponding to the single-peaked lightcurve is located at 1.38~cycles/day (17.44~h, Figure~\ref{fig:JL80}). Assuming a double-peaked lightcurve, the rotational period is 34.87~h. Such a slow rotation means that 2014~JL$_{80}$ is one of the few slow rotators detected from ground-based observations with Pluto/Charon and 2010~WG$_{9}$ \citep{Rabinowitz2013, Buie1997}. In Figure~\ref{fig:JL80}, we plot the corresponding lightcurve with a variability of 0.55~mag. Unfortunately, we do not have the U-shape of the curve, but the V-shape of the second minima is visible. It is important to point out that the lightcurve of 2014~JL$_{80}$ is very sparse and thus the rotational period estimate needs to be improve. In Figure~\ref{fig:JL80}, one can appreciate that there are several aliases of the main peak located at 2.37~cycles/day (10.11~h) and 0.77~cycles/day (31.03~h). In all cases, the rotation of 2014~JL$_{80}$ is very slow. For the purpose of the following work, we will use the main peak as rotational period estimate.   

A test using the full-width-at-half-maximum (FWHM) of the lightcurve's peaks can suggest if an object is potentially a close binary or an elongated object \citep{Thirouin2017b}. In fact, by estimating the FWHM of the peaks, we have shown that the FWHM of the peaks are different in case of single and contact binary objects using their lightcurve plotted as magnitude versus rotational phase. Unfortunately, because the lightcurve of 2014~JL$_{80}$ is sparse sampled, we are not able to provide an estimate for the U- and V-FWHM. But, based on the slow rotation of this object we can argue that tidal effects due to a companion have affected the rotational period of this object \citep{Thirouin2014}. In fact, several TNOs (Pluto-Charon \citep{Buie1997}, Sila-Nunam \citep{Grundy2012}, and maybe 2010~WG$_{9}$ \citep{Rabinowitz2013}) have rotational periods longer than a day and are synchronously locked binaries.

Following the procedure described in \citet{Rabinowitz2013}, and assuming that 2014~JL$_{80}$ is an equal-sized binary system, we can estimate the separation between the two components. With a primary and a secondary having comparable diameters of 150~km, a density of 1000~kg~m$^{-3}$, we calculate a $\sim$320~km separation. Such a separation is below the resolution of the \textit{Hubble Space Telescope} \citep{Noll2008}. 

Because of the large variability and the second minima shape, 2014~JL$_{80}$ is likely a contact binary or nearly contact binary and this case will be discussed in the following section.     
 
\subsubsection{2014~JO$_{80}$}

The absolute magnitude of 2014~JO$_{80}$ is 7.4~mag (MPC value), thus the diameter is 220/98~km using an albedo of 0.04/0.20.

\begin{figure*}
\includegraphics[width=9cm, angle=0]{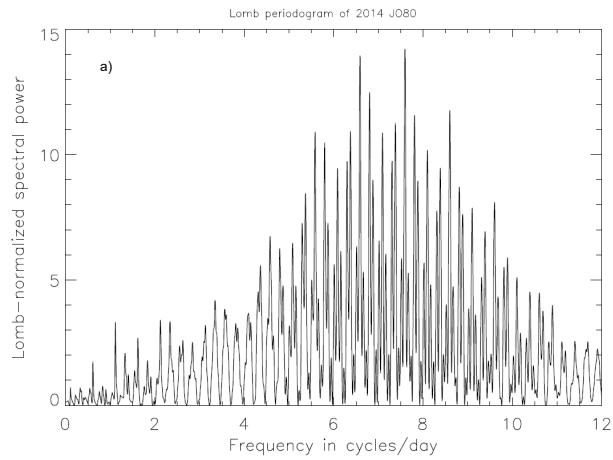}
\includegraphics[width=10cm, angle=0]{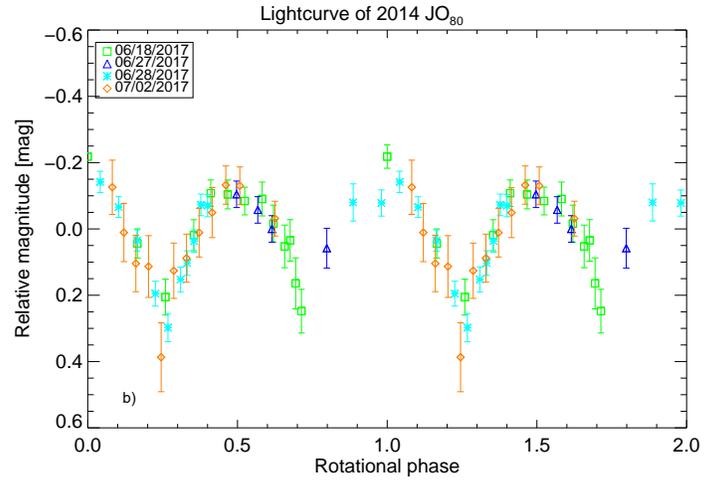}
\includegraphics[width=9cm, angle=0]{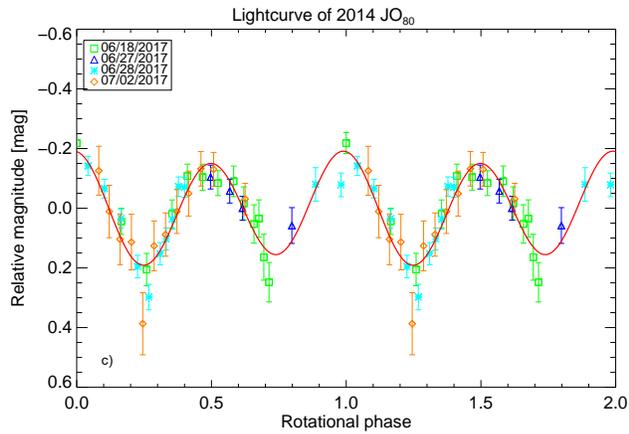}
\includegraphics[width=9cm, angle=0]{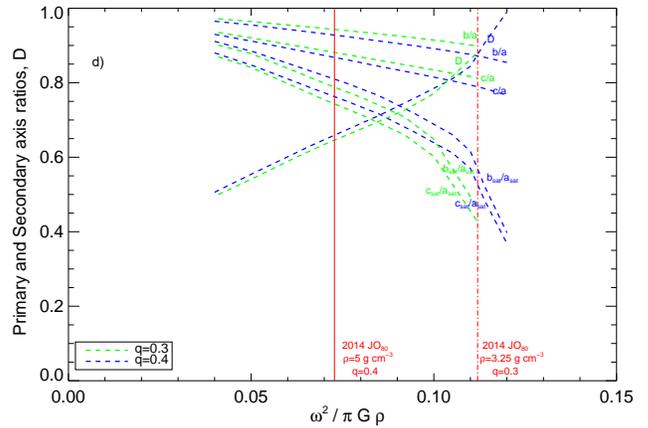}
\caption{\textit{Study of 2014~JO$_{80}$:} The highest peak of the Lomb periodogram favors a rotational period of 3.16~h (plot a)). However, based on the asymmetry of the lightcurve and the large amplitude, we favor the double-peaked option with a rotational period of 6.32~h (plot b)). A second order Fourier Series fit is not able to reproduce the V- and U-shape of the lightcurve, therefore, we propose that 2014~JO$_{80}$ is likely a contact binary (plot c)). In order to derive basic information about the system, we used \citet{Leone1984} work and summarized our results in the plot d).  }
\label{fig:JO80}
\end{figure*}

With four nights of data, the Lomb periodogram favors a peak at 7.59~cycles/day (3.16~h, Figure~\ref{fig:JO80}). There are also aliases of the main peak located at 6.58~cycles/day (3.65~h) and 8.58~cycles/day (2.80~h). All techniques used to confirm the periodicity favors the main peak and thus we are using this value for our work. Because of the large amplitude and because of an asymmetry of about 0.1~mag between the peaks, we favor the double-peaked lightcurve with a rotational period of 6.32~h. The variability is 0.60$\pm$0.05~mag (Figure~\ref{fig:JO80}). 

Following \citet{Thirouin2017b}, we estimate the U-FWHM at 0.38 and 0.37, and the V-FWHM at 0.16. Therefore, 2014~JO$_{80}$ presents the same characteristic as the likely contact binaries reported in \citet{Thirouin2017b}.

\subsubsection{2014~JQ$_{80}$}

Based on the MPC estimate, the absolute magnitude of 2014~JQ$_{80}$ is 7.0~mag, suggesting a diameter of 265/118~km with an albedo of 0.04/0.20. 

\begin{figure*}
\includegraphics[width=9cm, angle=0]{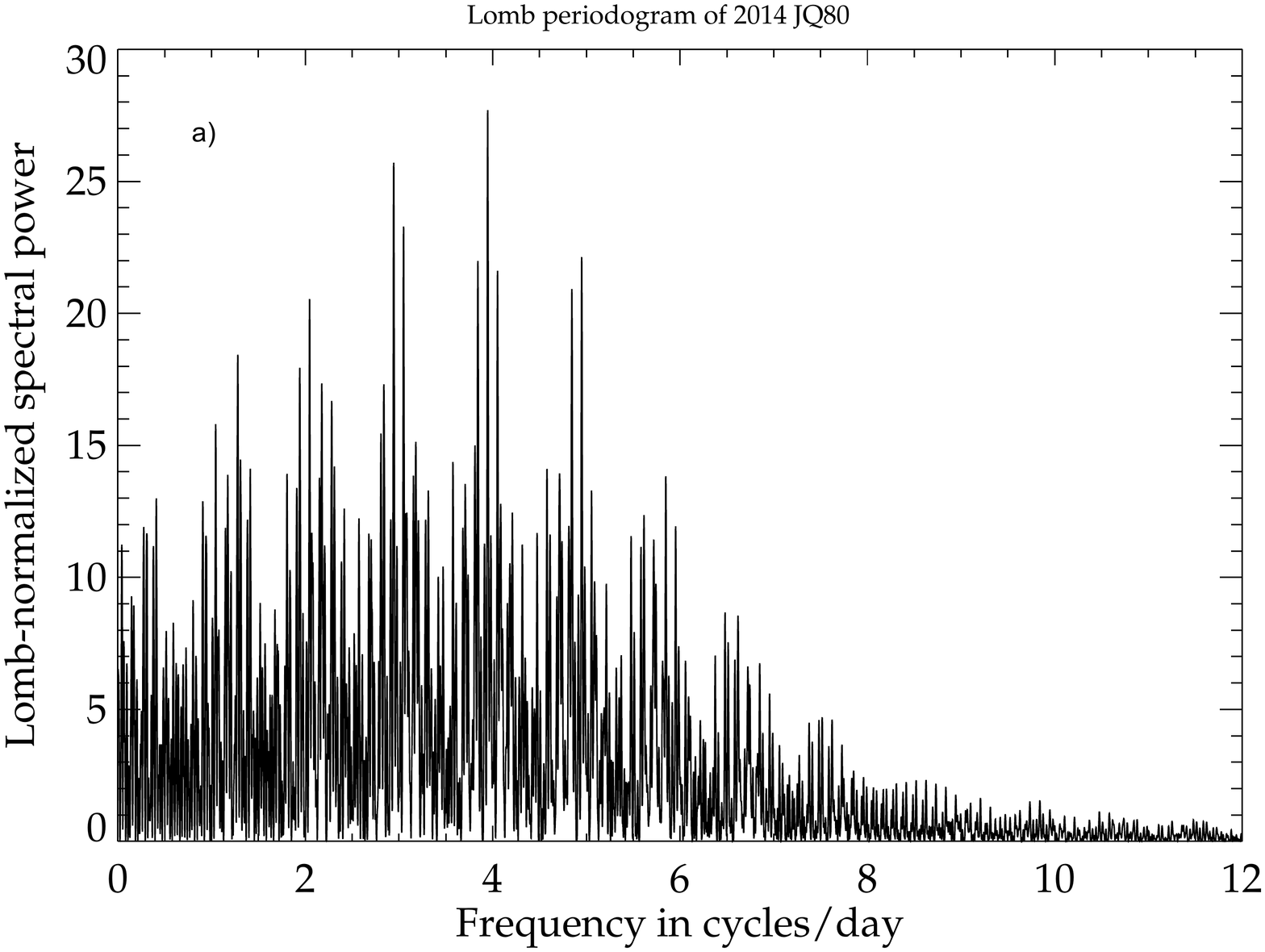}
\includegraphics[width=10cm, angle=0]{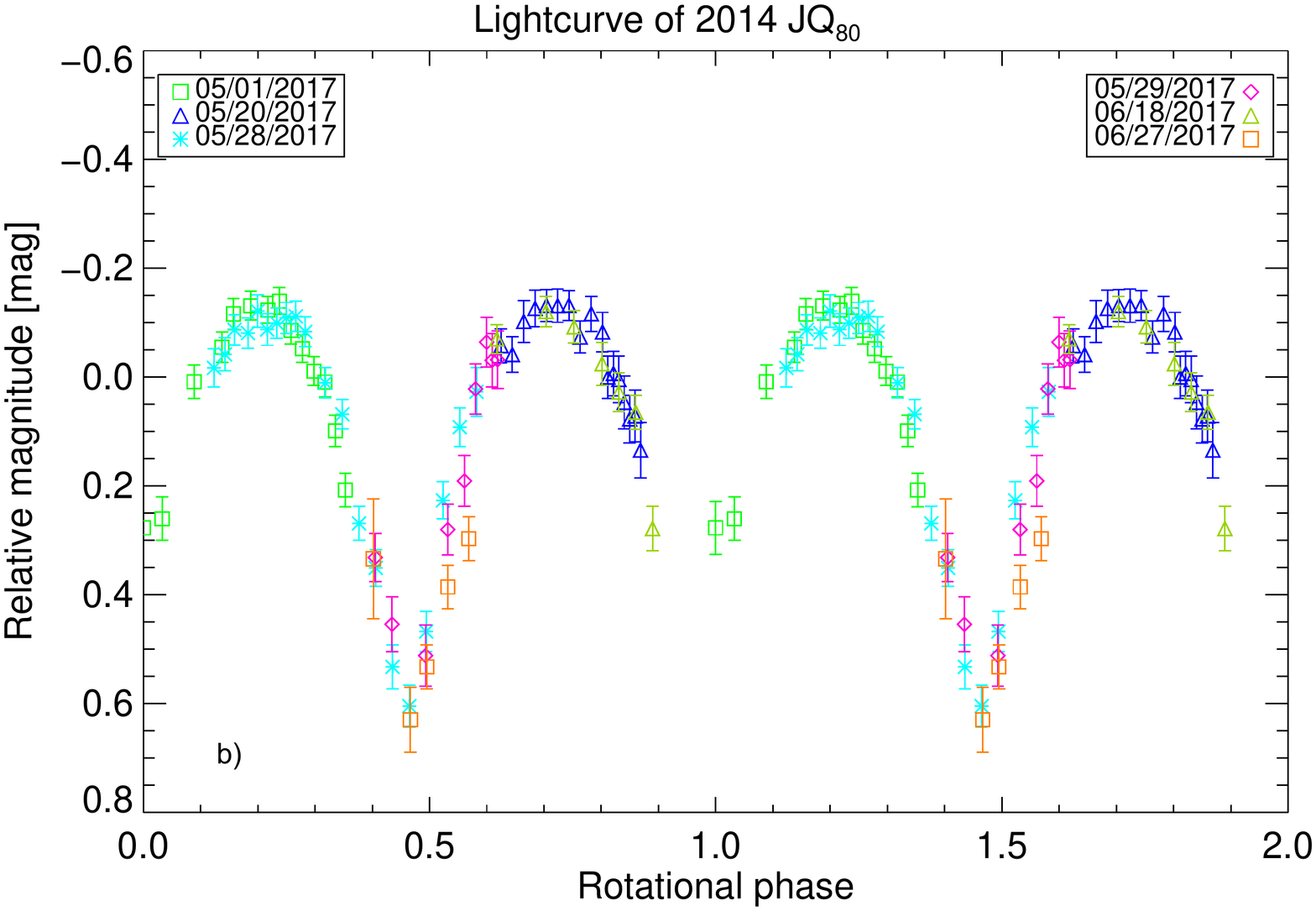} 
\includegraphics[width=9cm, angle=0]{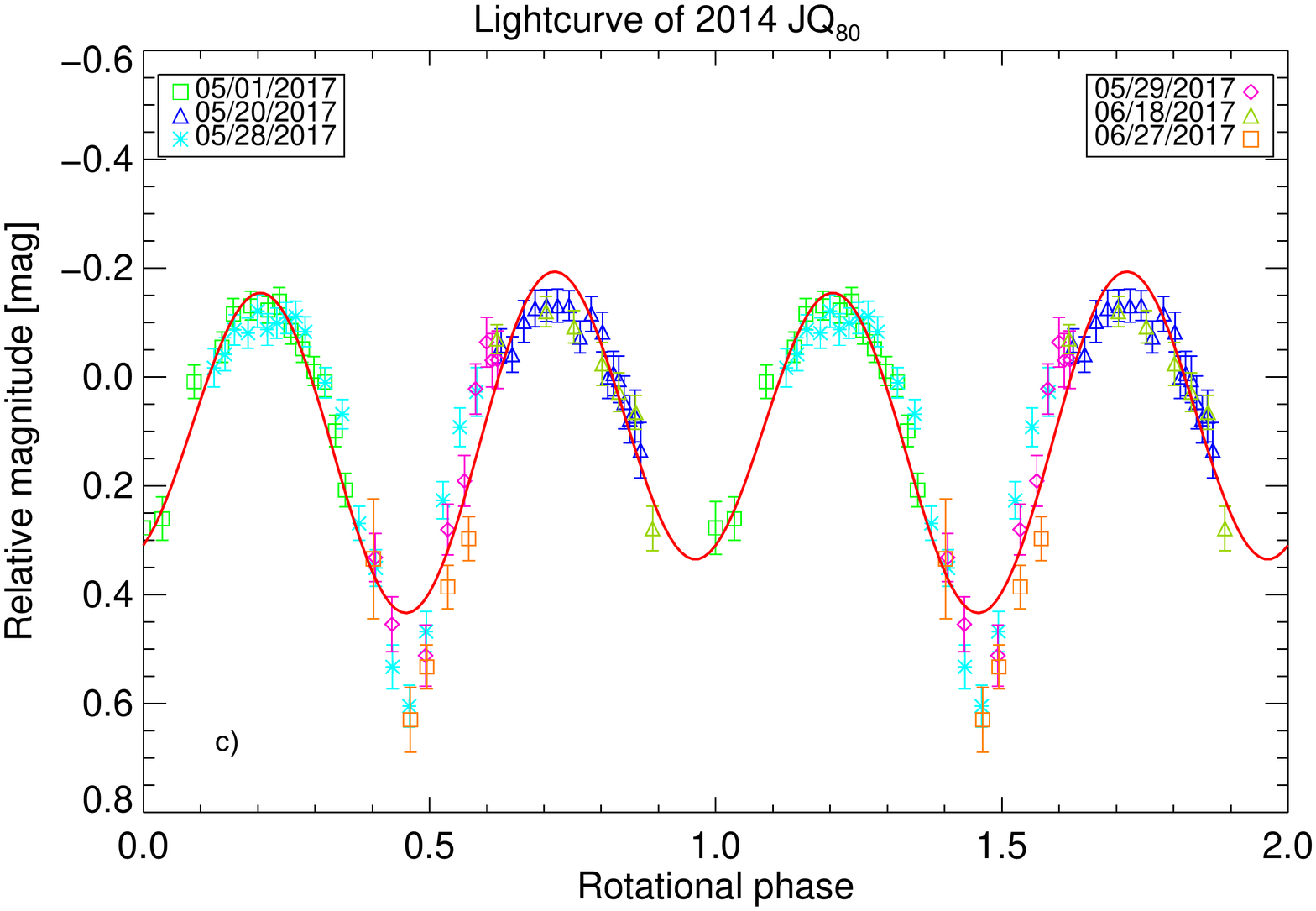}
\includegraphics[width=9cm, angle=0]{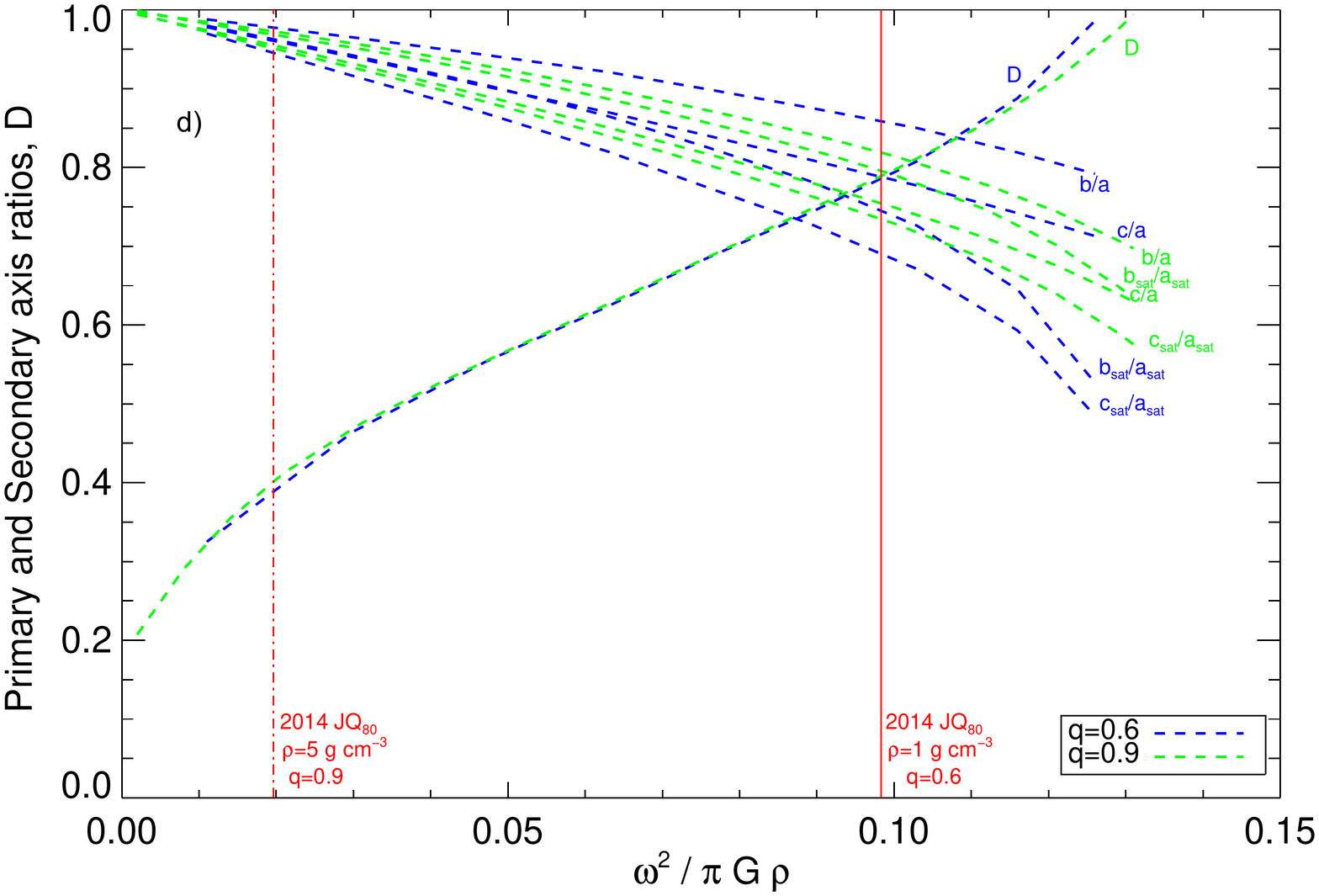}
\caption{\textit{Study of 2014~JQ$_{80}$:} Based on the Lomb periodogram study, we estimate that 2014~JQ$_{80}$  has a double-peaked lightcurve with a rotational period of 12.16~h (plot a)). Because the lightcurve displays a large variability, and because of the V-/U-shapes of the minima/maxima, we suggest that this object is a contact binary (plot b), and c)). The plot d) was used to derive the basic information of the system, assuming a contact binary nature.  }
\label{fig:JQ80}
\end{figure*}

With a total of 83 images obtained over two months in 2017, we report the first and unique lightcurve of 2014~JQ$_{80}$. The Lomb periodogram, and the PDM technique favor a rotational period of 3.97~cycles/day (6.08~h, Figure~\ref{fig:JQ80}), but there two aliases of this main peak with lower confidence levels at 2.95~cycles/day (8.13~h) and 4.94~cycles/day (4.86~h). Following, we will use the main peak with the highest confidence level as the rotational periodicity. Because of the large amplitude of this object, we favor the double-peaked lightcurve with a rotational period of 12.16~h. The lightcurve amplitude is 0.76$\pm$0.04~mag (Figure~\ref{fig:JQ80}).

With the U-FWHM of 0.34 and 0.35, and the V-FHWM of 0.16 and 0.13, 2014~JQ$_{80}$ meets the criteria mentioned in \citet{Thirouin2017b} for likely contact binaries.   \\

\begin{figure*}
\includegraphics[width=12cm, angle=0]{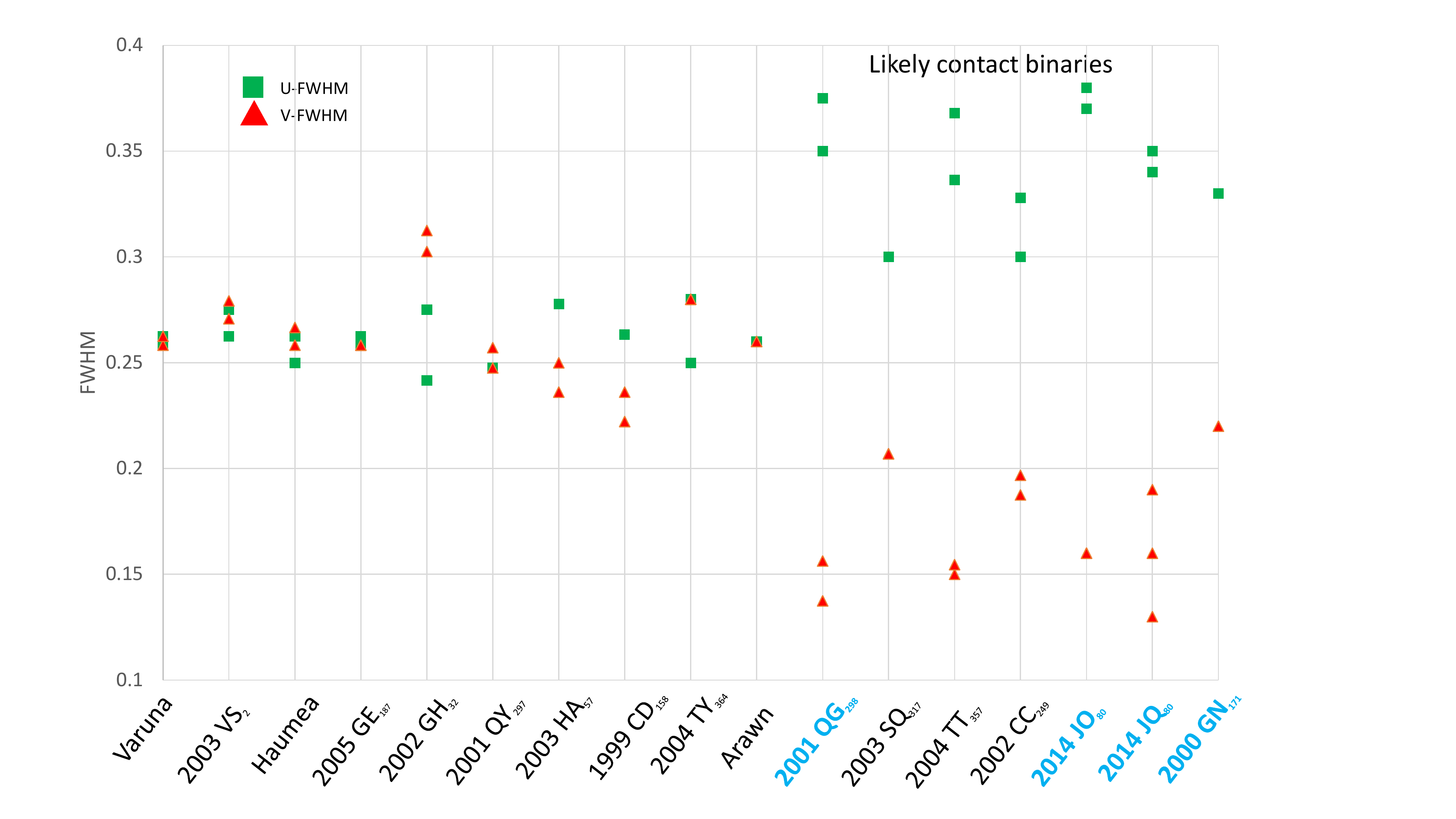}
\caption{\textit{Full width at half maximum (FWHM) of single objects, resolved binaries and (likely/confirmed)
contact binaries}: A complete description of this plot is available in \citet{Thirouin2017b}. The (likely/confirmed) contact binaries present a U-FWHM$\geq$0.30, and a V-FWHM $\leq$0.22. Likely contact binary in the Plutino population have their name highlighted in light blue. This study made use of lightcurves plotted as magnitude versus rotational phase.   }
\label{fig:FWHM}
\end{figure*}

Our study regarding the U-/V-FWHM is summarized in Figure~\ref{fig:FWHM}. A complete description of this plot is available in \citet{Thirouin2017b}.

\subsection{Partial lightcurves}

\subsubsection{1995~HM$_{5}$}

\begin{figure*}
\includegraphics[width=9cm, angle=0]{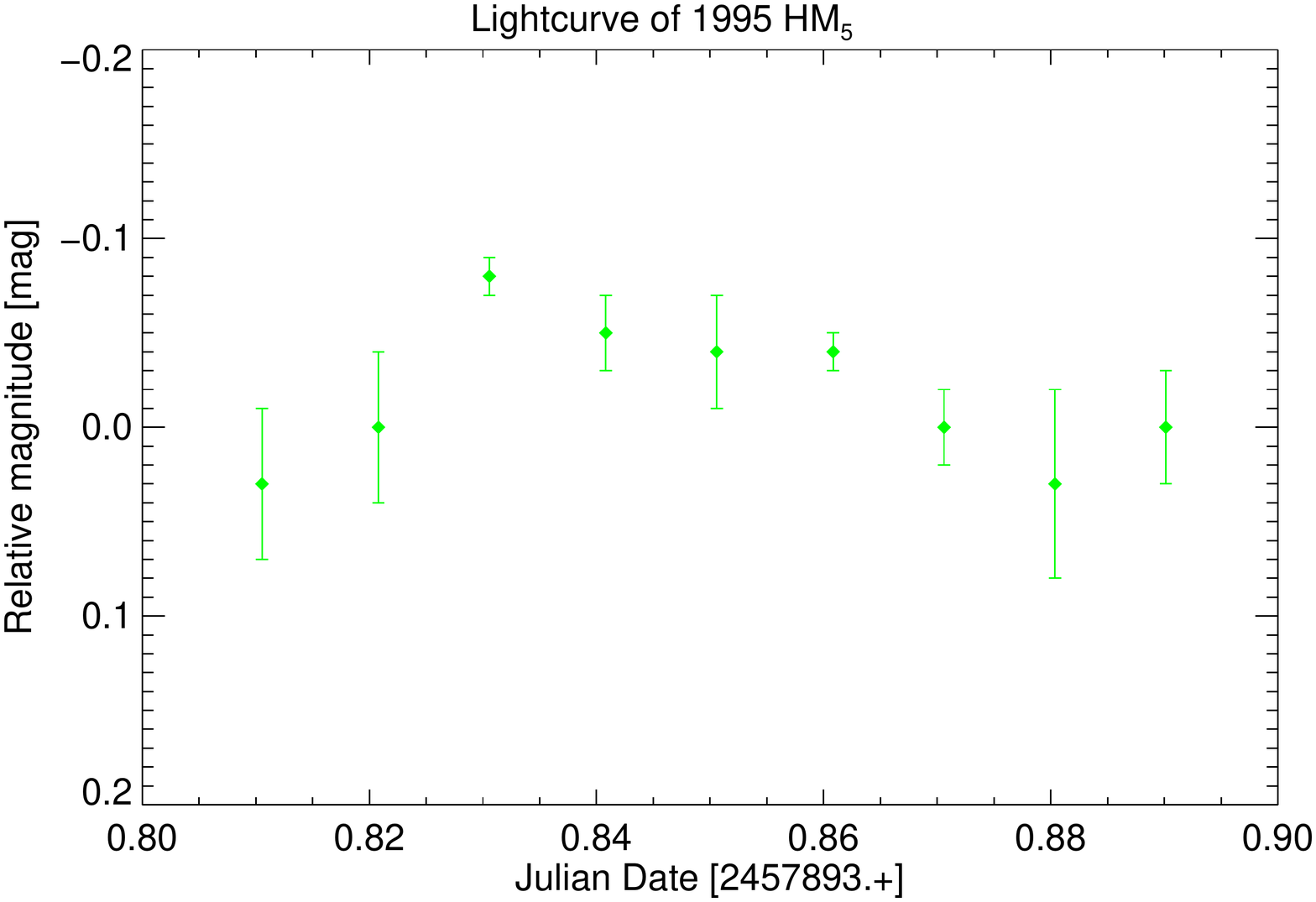}
\includegraphics[width=9cm, angle=0]{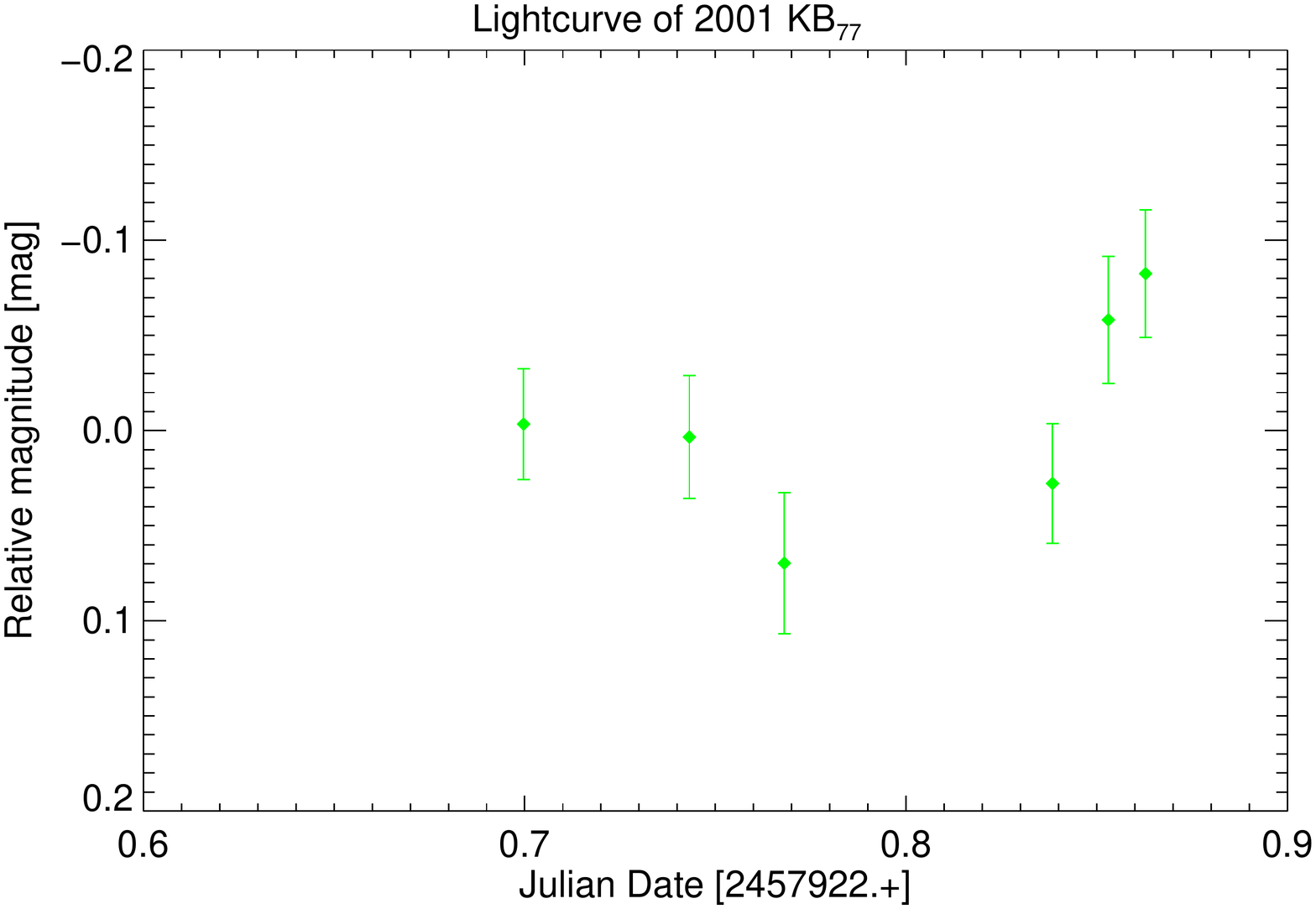}
\includegraphics[width=9cm, angle=0]{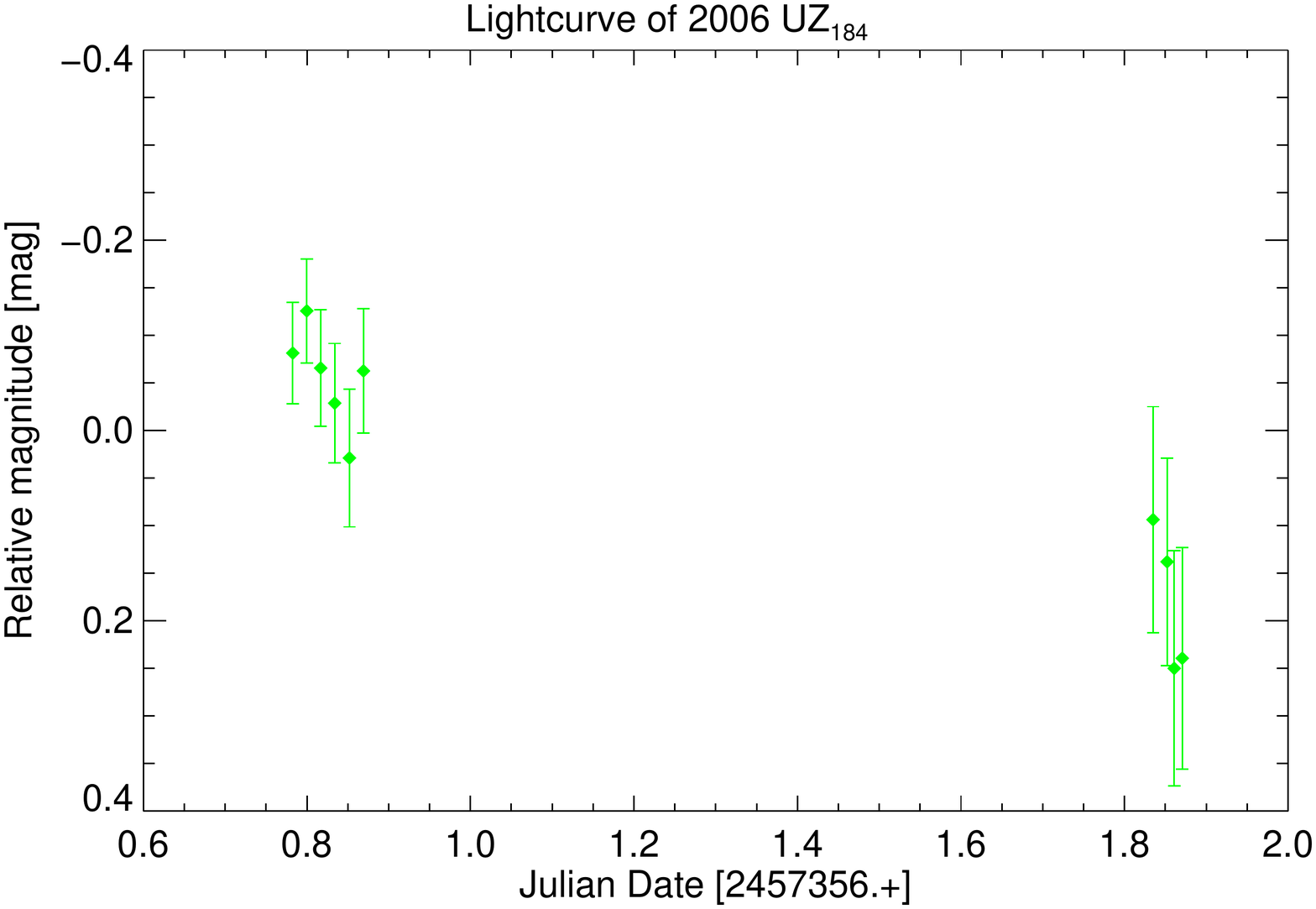}
\includegraphics[width=9cm, angle=0]{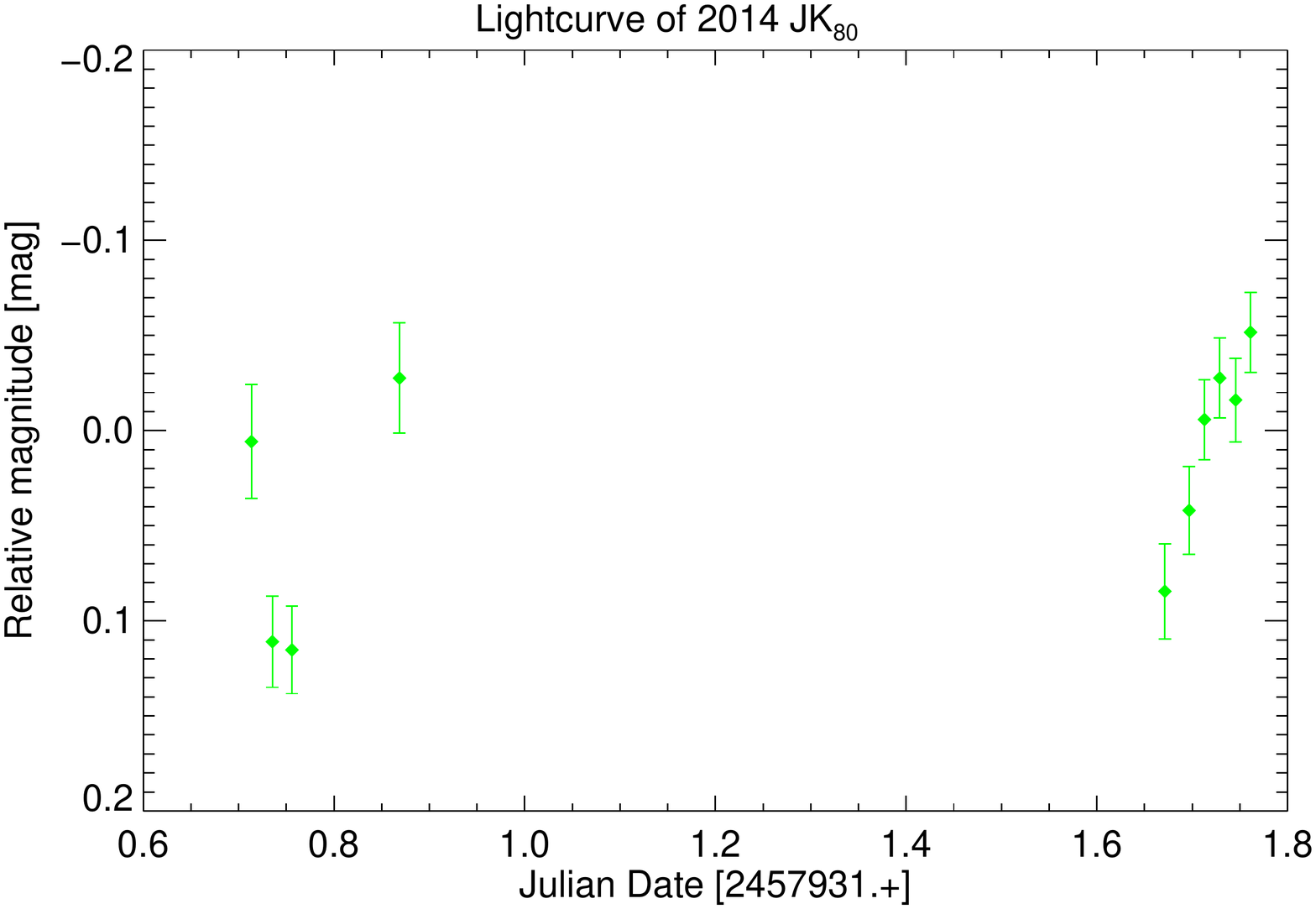}
\includegraphics[width=9cm, angle=0]{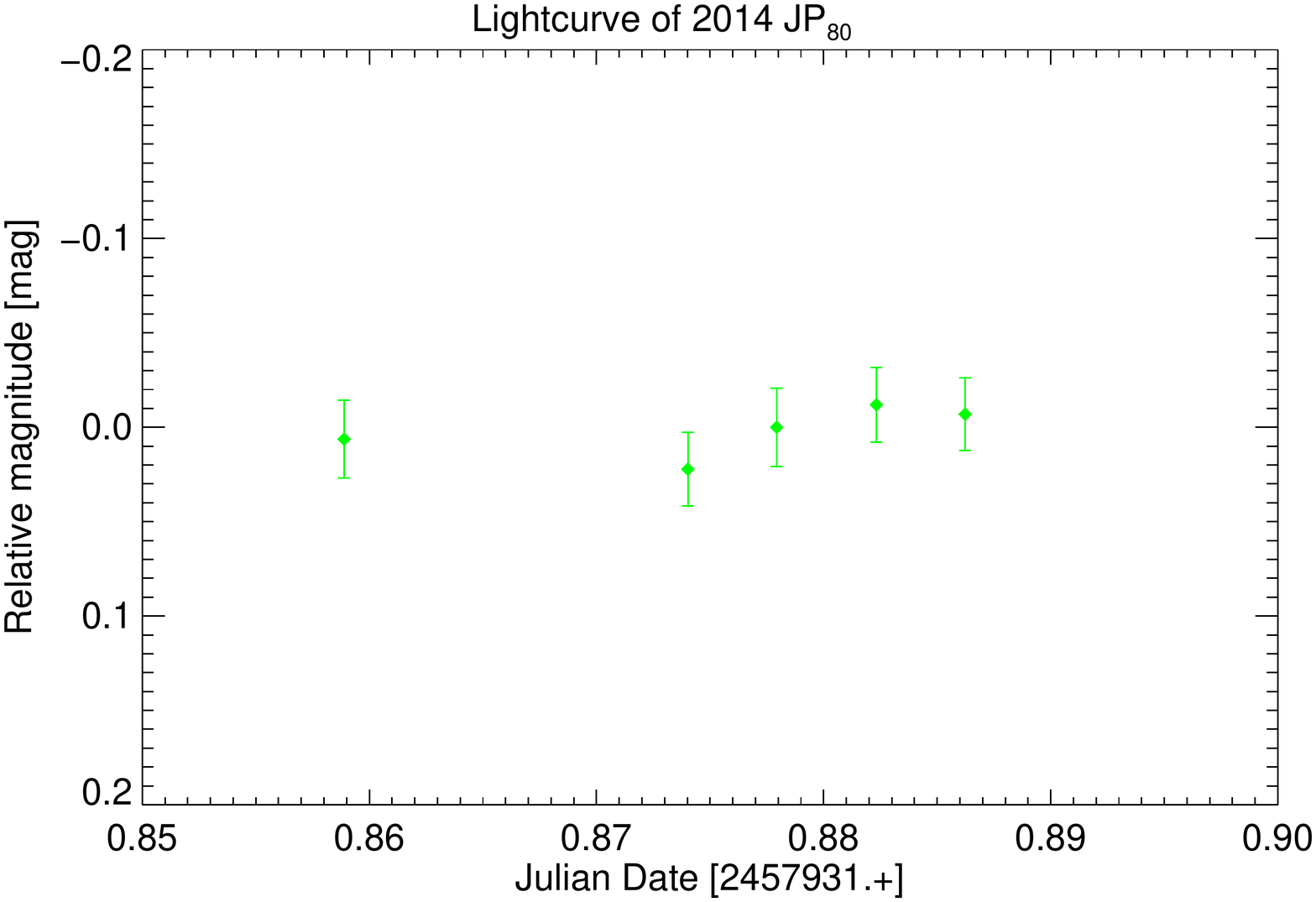}
\includegraphics[width=9cm, angle=0]{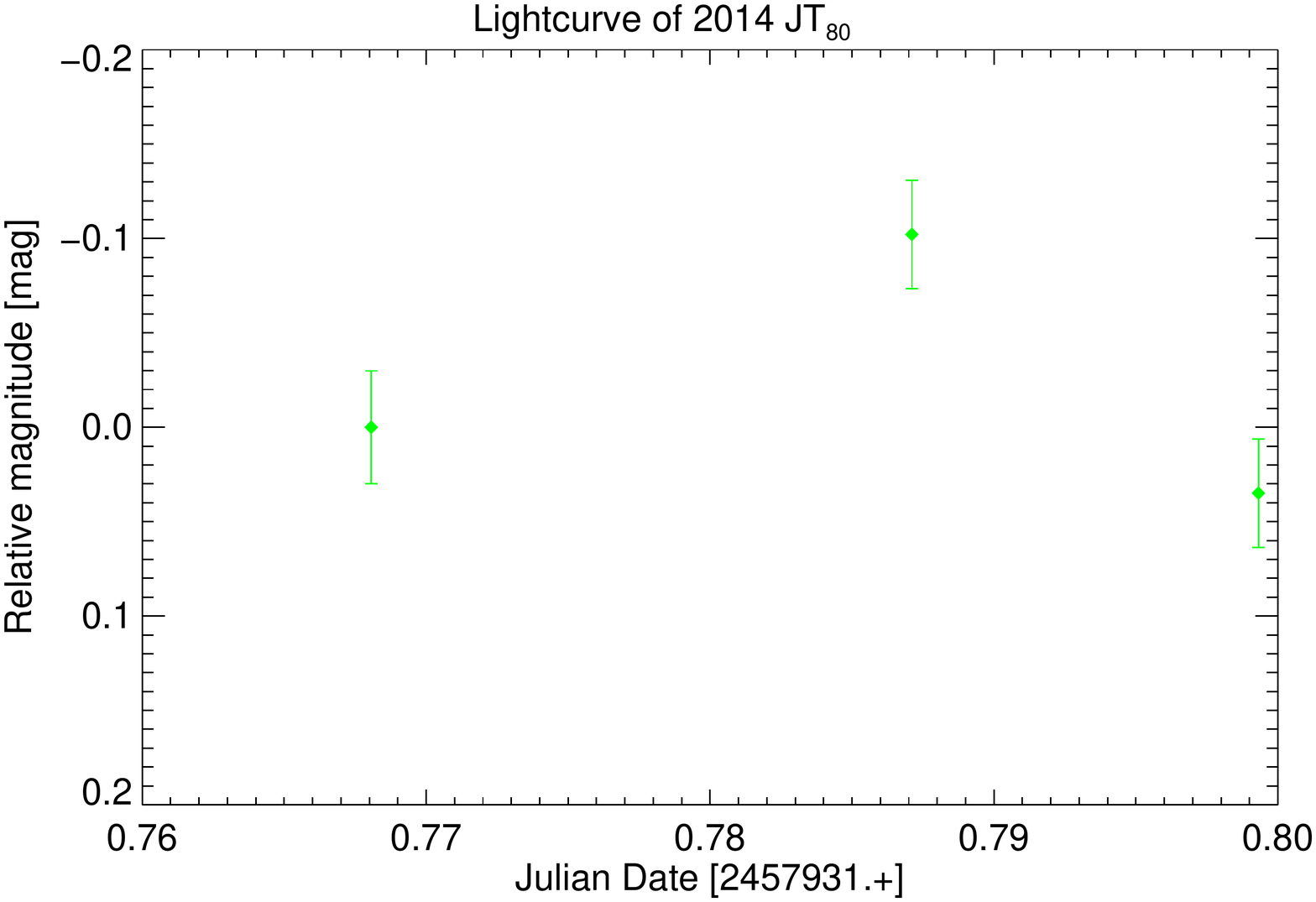} 
\caption{\textit{Partial lightcurves of several Plutinos:} we report relative magnitude versus Julian Date (no light-time correction) for nine objects.  }
\label{fig:LCnocb}
\end{figure*}

1995~HM$_{5}$ is the smallest object in our sample. With an absolute magnitude of 7.7~mag, its diameter is between 86 and 192~km assuming a geometric albedo of 0.04 and 0.2, respectively. We report about 0.6~h of observations obtained in May 2017. Our few observations show a minimum variability of $\sim$0.1~mag (Figure~\ref{fig:LCnocb}). Based on our data, no rotational period is derived.     

\subsubsection{(469362) 2001~KB$_{77}$}

We report 6 images of 2001~KB$_{77}$ obtained on UT 18 June. In about 4~h, 2001~KB$_{77}$ displays a variability of 0.15~mag (Figure~\ref{fig:LCnocb}). To our knowledge, there is no bibliographic reference to compare our result with. No rotational period is derived from our observations, we can only constrain the period to be longer than 4~h. 

\subsubsection{2006~UZ$_{84}$}

We observed 2006~UZ$_{184}$ over two consecutive nights in November-December 2015. Unfortunately, due to bad weather conditions, only 10 images were usable. Over two nights, this object seems to have a variability of about 0.2~mag, and no period is derived (Figure~\ref{fig:LCnocb}).   

\subsubsection{2014~JK$_{80}$}

2014~JK$_{80}$ was studied over two consecutive nights at the end of June 2017. Data from the first night are not showing an obvious sign of variability, but the second night shows a variability of about 0.17~mag (Figure~\ref{fig:LCnocb}). 
 
\subsubsection{2014~JP$_{80}$}

2014~JP$_{80}$, with H=4.9~mag, a diameter of 698/311~km considering an albedo of 0.04/0.2 is the largest object here. With only 5 images obtained over about 0.5~h, it is difficult to provide any reliable conclusion regarding 2014~JP$_{80}$. The partial lightcurve shows a variability of $\sim$0.1~mag (Figure~\ref{fig:LCnocb}). 

\subsubsection{2014~JT$_{80}$}

We have five images of 2014~JT$_{80}$ obtained over about 1~h in June 27, but only 3 images were usable. The variabilty is about 0.1~mag (Figure~\ref{fig:LCnocb}). 

\subsubsection{2014~KC$_{102}$}

We have two nights of data for 2014~KC$_{102}$. The second night shows the maximum of the curve with a variability of around 0.2~mag over 4.5~h of observations (Figure~\ref{fig:LCnocb}). Therefore, considering a double-peaked lightcurve, we suggest a rotational period of about 9~h, but more data are needed for a better estimate.  

\subsubsection{2014~KX$_{101}$}

We report two consecutive nights of observations for 2014~KX$_{101}$. Both nights are showing a variability of about 0.2~mag (Figure~\ref{fig:LCnocb}). Unfortunately, no rotational period estimate is derived. 

\subsubsection{2015~BA$_{519}$}

2015~BA$_{519}$ was observed during only one night on UT May 1. In a couple of hours of observations, a variability of $\sim$0.16~mag is noticed. The partial lightcurve plotted in Figure~\ref{fig:LCnocb} presents one maximum and one minimum, so the rotational period of this object should be approximately 8~h (assuming a double-peaked lightcurve). No additional information about this object has been found in the literature.
 

\section{Analysis}
\label{sec:analysis}

Next, we consider a contact binary configuration and a single very elongated object option to explain the high lightcurve variability of 2014~JL$_{80}$, 2014~JO$_{80}$, and 2014~JQ$_{80}$.

 \subsection{Roche system}

The large lightcurve amplitude of 2014~JL$_{80}$, 2014~JO$_{80}$, and 2014~JQ$_{80}$ are best explained if these objects are contact binary systems. Following, we constrain basic information about the systems \citet{Leone1984}. We estimate the mass ratio and the density using the Roche sequences (Figure~\ref{fig:Roche}). Following \citet{Leone1984} assumptions, we derive min and max cases. The density range in \citet{Leone1984} is limited to $\rho$=1000~kg~m$^{-3}$-5000~kg~m$^{-3}$, but lower densities are possible for the TNOs.

\subsubsection{2014~JL$_{80}$}

For 2014~JL$_{80}$, based on \citet{Leone1984} we estimate that q$_{min}$$\sim$q$_{max}$$\sim$0.5, whereas the density range is $\rho_{min}$=1000~kg~m$^{-3}$, $\rho_{max}$=5000~kg~m$^{-3}$. 

Considering 2014~JL$_{80}$ as a binary with a mass ratio of 0.5, and a density of 1000~kg~m$^{-3}$, the axis ratios of the primary are: b/a=0.99, c/a=0.98 (a=71/31~km, b=71/31~km, and c=70/31~km assuming an albedo of 0.04/0.2), the axis ratios of the secondary are: b$_{sat}$/a$_{sat}$=0.98, c$_{sat}$/a$_{sat}$=0.97 (a$_{sat}$=57/26~km, b$_{sat}$=56/26~km, and c$_{sat}$=55/25~km with an albedo of 0.04/0.2). The parameter D is 0.34, thus the distance between bodies is 378/169~km (albedo of 0.04/0.2).

\subsubsection{2014~JO$_{80}$}

For 2014~JO$_{80}$, we have: i) a system  with q$_{min}$=0.3 and $\rho_{min}$=3.25~g~cm$^{-3}$ or ii) a system q$_{max}$=0.42 and $\rho_{max}$=5~g~cm$^{-3}$. Based on the lightcurve amplitude, the uncertainty for the mass ratio is $\pm$0.07. 
If 2014~JO$_{80}$ is a binary system with a mass ratio of 0.3, and a density of 3.25~g~cm$^{-3}$, we derive the axis ratios of the primary: b/a=0.90, c/a=0.81 (a=75/33~km, b=68/30~km, and c=61/27~km assuming an albedo of 0.04/0.2), the axis ratios of the secondary: b$_{sat}$/a$_{sat}$=0.46, c$_{sat}$/a$_{sat}$=0.43 (a$_{sat}$=79/36~km, b$_{sat}$=36/16~km, and c$_{sat}$=34/15~km assuming an albedo of 0.04/0.2). The separation between the components is 175/78~km with an albedo of 0.04/0.2, assuming D=0.88.

\subsubsection{2014~JQ$_{80}$}

For 2014~JQ$_{80}$, we compute two extreme options: i) a system with q$_{min}$=0.57 and $\rho_{min}$=1~g~cm$^{-3}$ or ii) a system with q$_{max}$=0.92 and $\rho_{max}$=5~g~cm$^{-3}$. Following, we will consider conservative mass ratio of q$_{min}$=0.6. 

If 2014~JQ$_{80}$ is a binary with q=0.6, and $\rho$=1~g~cm$^{-3}$, we obtain for the primary: b/a=0.86, c/a=0.79 (a=84/37~km, b=72/32~km, and c=66/29~km assuming an albedo of 0.04/0.2), for the secondary: b$_{sat}$/a$_{sat}$=0.75, c$_{sat}$/a$_{sat}$=0.69 (a$_{sat}$=78/35~km, b$_{sat}$=59/26~km, and c$_{sat}$=54/24~km assuming an albedo of 0.04/0.2). The separation between the components is 208/92~km with an albedo of 0.04/0.2 and D=0.78. \\

The typical density in the Kuiper belt is $\sim$1~g~cm$^{-3}$, with the exceptions of a few denser bigger objects \citep{Thirouin2016, Vilenius2014, Brown2013, Grundy2012, Sheppard2008}. Therefore, previously we do not derive axis ratios and separation assuming a high density $\rho_{max}$=5~g~cm$^{-3}$. It is important to point out that a careful modelling of these systems using several lightcurves obtained a different epochs will be necessary to derive accurate basic parameters for all of them.

\subsection{Jacobi ellipsoid}

The lightcurves of 2014~JL$_{80}$, 2014~JO$_{80}$, and 2014~JQ$_{80}$ are not reproduced by a second order Fourier series fit because of their U-/V-shapes. Therefore, we favor the option of contact binary to explain these lightcurves, but if these objects are single Jacobi ellipsoids, we can constrain their elongations, and densities \citep{Thirouin2017, Thirouin2017b}.
 
\subsubsection{2014~JL$_{80}$}

In the case of 2014~JL$_{80}$, considering a viewing angle of 90$^\circ$, we find a/b=1.67, and c/a=0.44 \citep{Chandrasekhar1987}. So, we compute: a=213~km (a=95~km), b=128~km (b=57~km), and c=94~km (c=42~km) for an albedo of 0.04 (0.20) and an equatorial view. Using a viewing angle of 60$^\circ$, we derive an axis ratio a/b$>$2.31. As ellipsoids with a/b$>$2.31 are unstable to rotational fission, the viewing angle of 2014~JL$_{80}$ must be larger than 62.5$^\circ$ \citep{Jeans1919}. With an equatorial view, the density is $\rho$$\geq$0.04~g~cm$^{-3}$. 
The very long rotation of 2014~JL$_{80}$ likely means that this object cannot be a Jacobi ellipsoid which generally are fast rotators and have high angular momentum \citep{Sheppard2008}. 

\subsubsection{2014~JO$_{80}$}

Assuming an equatorial view, we compute a/b=1.72, and c/a=0.42. Therefore, the axis are: a=170~km (a=85~km), b=99~km (b=49~km), and c=71~km (c=36~km) for an albedo of 0.04 (0.20) and an equatorial view. As for 2014~JL$_{80}$, a viewing angle of 60$^\circ$ suggests that 2014~JO$_{80}$ is unstable to rotational fission. We estimate that the aspect angle is between 65$^\circ$ and 90$^\circ$. Using an equatorial view, the density is $\rho$$\geq$1100~kg~m$^{-3}$. 
The very short period of 2014~JO$_{80}$ makes a Jacobi type object likely because of its high angular momentum.

\subsubsection{2014~JQ$_{80}$}

With an equatorial view for 2014~JQ$_{80}$, we obtain a/b=2, and c/a=0.38. Therefore, we compute: a=248~km (a=111~km), b=124~km (b=55~km), and c=94~km (c=42~km) for an albedo of 0.04 (0.20) and an equatorial view. 2014~JQ$_{80}$ is stable to rotational fission if its viewing angle is larger than 74$^\circ$. Assuming $\xi$=90$^\circ$, the density is $\rho$$\geq$0.32~g~cm$^{-3}$. \\

In conclusion, to explain the morphology of the lightcurves of 2014~JL$_{80}$, 2014~JO$_{80}$, and 2014~JQ$_{80}$, we showed details of a single elongated object or close binary configuration. However, because of the V-/U-shape, and the extreme variability, we favor the contact binary explanation. Future observations are required to infer the nature of 2014~JL$_{80}$, 2014~JO$_{80}$, and 2014~JQ$_{80}$. It is also important to point out that we used the rotational periods favored by the main peaks (i.e. peak with the highest confidence level), however more data can be useful to favor such peaks and not some of the aliases.  

\section{Percentage of contact binaries}

\subsection{Status of the Plutino lightcurve studies}

In Table~\ref{Summary_photo}, and Table~\ref{Tab:photoplutino} are summarized the short-term variability studies of Plutinos published in this work, and in the literature (respectively). In total, 39 Plutinos\footnote{To date, 225 Plutinos are known. So, based on published results, only $\sim$17$\%$ of the Plutinos have been observed for lightcurves. } have been observed for lightcurves: 12 Plutinos are reported in this paper and 27 in the literature. However, only 21 objects have a rotational period (secure result and tentative), the rest have only lightcurve amplitude constraints.  

Based on Table~\ref{Tab:photoplutino}, one can appreciate that most of the Plutinos display a low lightcurve amplitude with a mean value of $\sim$0.13~mag (Figure~\ref{fig:all}). But, several objects are showing larger variability: Pluto, Arawn (1994~JR$_{1}$), 1995~QY$_{9}$, 2000~GN$_{171}$, and 2001~QG$_{298}$. The NASA's New Horizons flyby of Pluto confirmed that the lightcurve amplitude is due to a strong albedo contrast on Pluto's surface, whereas 2001~QG$_{298}$ is a contact binary \citep{Buratti2017, Sheppard2004}. Arawn, also observed by New Horizons at a very high phase angle, displays an amplitude of 0.58~mag\footnote{\citet{Porter2016} reported an amplitude of 0.8~mag, being the difference between the highest and lowest point of the curve.} based on the fit reported in \citet{Porter2016}. We estimate the V-FWHM and the U-FWNM of Arawn and find that the U-FWHM and the V-FWHM are about the same ($\sim$0.26). Therefore, following the criteria reported in \citet{Thirouin2017b}, Arawn is likely not a contact binary, and thus its large amplitude is due to an elongated shape. However, the Arawn observations were performed at a phase angle up to $\sim$59$^\circ$ whereas the typical ground-based TNO lightcurves are obtained at a phase angle up to $\sim$2$^\circ$. With such a high phase angle compared to Earth observations, it is difficult to directly compare the lightcurve morphologies as shadowing effect is not negligible. Therefore, the direct comparison of ground- and space-based lightcurves can be wrong. In the case of 1995~QY$_{9}$, only one lightcurve is reported in the literature \citep{Romanishin1999}. The lightcurve amplitude is about 0.6~mag. Unfortunately, no further study of this object can confirm or not such finding, and we cannot use the V-/U-FWHM technique as the photometry is not available and the lightcurve plotted is not phased folded. 

The lightcurve of 2000~GN$_{171}$ displays a large variability of $\sim$0.6~mag \citep{Sheppard2002, Rabinowitz2007}. It has been suggested that the lightcurve of 2000~GN$_{171}$ is due to an elongated shape, but the option of a contact binary system was not excluded \citep{Sheppard2002}. Later, based on lightcurve modeling \citet{Lacerda2007} suggested that the Roche and Jacobi solution are equivalent, and thus could not favor any option regarding the nature of this object/system. Using \citet{Sheppard2002} lightcurve, we estimate an U-FWHM of 0.33 and a V-FHWM of 0.22 and 0.19, therefore according to our study reported in \citet{Thirouin2017b}, 2000~GN$_{171}$ is likely a contact binary (Figure~\ref{fig:FWHM}). The most recent lightcurve of 2000~GN$_{171}$ was obtained in 2007 \citep{Dotto2008}. They obtained a rotational period consistent with \citet{Sheppard2002} result, and they argue that the lightcurve amplitude is also similar. However, based on their Figure~3, the lightcurve amplitude is smaller than reported. Their photometry is not available online, but using their plot, the amplitude seems to be around 0.53~mag. In conclusion, there is a potential change in the lightcurve amplitude and thus potentially a change in the system orientation between the first and last lightcurve obtained in 2001 and 2007. A careful modeling of those lightcurves may infer the nature of the system/object.  \\

In conclusion, 2000~GN$_{171}$ is likely a contact binary and a change in the lightcurve amplitude is potentially observable. In the case of 1995~QY$_{9}$, a new lightcurve will be useful to constrain the nature of this object. For the purpose of our following study, we will assume that this object is a single elongated one, as it is the case for Arawn.

\subsection{Fraction of contact binaries in the Plutino population}

With the literature and this work, we have evidence that 5 likely/confirmed contact binaries are in the Plutino population, not including the large amplitude objects 1995~QY$_{9}$ and Arawn as we assume they are elongated but not contact binary. We want to point out that these 5 likely/confirmed contact binaries are based on the literature data and the objects observed for this work (i.e. 5 out of 39 objects). Such a number seems to indicate that the Plutino population may have a large reservoir of contact binaries compared to the other populations \citep{Thirouin2017b, Thirouin2017, Lacerda2014, Sheppard2004}. It is important to point out that a threshold of 0.9~mag has been used in the literature to identify a contact binary, but a non-equal-sized close system may never pass such a threshold even if observed equator-on. Therefore, in the next paragraphs we will use several lightcurve amplitude limits. We also assume that the partial lightcurves reported in this work are due to non-contact binary objects. However, some of them are maybe contact binary with very long rotational period and thus their identification is not possible based on our data. Therefore, the fractions of contact binaries reported below are lower limits. 

Following the procedure detailed in \citet{Sheppard2004}, we calculate the percentage of close binaries in the Plutino population. 

In first approximation, we assume that the lightcurve amplitude of an object with axes a$>$b, and b=c varies following: 
\begin{equation}
\Delta_m = 2.5 \log \left(\frac{1+\tan \theta}{(b/a)+\tan \theta}\right) 
\label{eq:frac1}
\end{equation}
$\theta$ is the angle of the object's pole relative to the perpendicular of the sight line. An amplitude of 0.9~mag is reached at $\theta$=10$^\circ$ for an object with an elongation of a/b=3. Therefore, the probability to observe an object from a random distribution within 10$^\circ$ from the sight line is P($\theta$$\leq$10$^\circ$)=0.17. With the same approach, we estimate that an amplitude of 0.5~mag is reached at $\theta$=36$^\circ$, whereas an amplitude of 0.6~mag is reached at $\theta$=27$^\circ$, and an amplitude of 0.7~mag is reached at $\theta$=20$^\circ$. In our sample of 12 objects we have one TNO with a $\Delta_m$$\geq$0.7~mag. So the detection of one TNO with $\Delta_m$$\geq$0.7mag implies that the abundance of similar objects is: f($\Delta_m$$\geq$0.7~mag)$\sim$1/(12*P($\theta$$\leq$20$^\circ$)) $\sim$25$\%$. Similarly, we estimate: f($\Delta_m$$\geq$0.6mag)$\sim$2/(12*P($\theta$$\leq$27$^\circ$)) $\sim$37$\%$, and f($\Delta_m$$\geq$0.5~mag)$\sim$3/(12*P($\theta$$\leq$36$^\circ$)) $\sim$42$\%$. Using our sample and the literature, we calculate: f($\Delta_m$$\geq$0.7~mag) $\sim$28$\%$, f($\Delta_m$$\geq$0.6~mag) $\sim$42$\%$, and f($\Delta_m$$\geq$0.5~mag) $\sim$40$\%$. Theses estimates do not include 1995~QY$_{9}$ and Arawn as we assume they are elongated objects.   \\
 
In a second approximation, and assuming a triaxial Jacobi with a$\geq$b=c, the lightcurve variability is \citep{Sheppard2004}: 
\begin{equation}
\Delta_m = 2.5 \log \left(\frac{a}{b}\right) - 1.25 \log \left(\left( \left(\frac{a}{b}\right)^2 -1 \right) \sin^2 \theta +1\right)
\label{eq:frac2}
\end{equation}
With the same approach as previously, we find for our sample: f($\Delta_m$$\geq$0.7~mag) $\sim$19$\%$, f($\Delta_m$$\geq$0.6~mag) $\sim$34$\%$, f($\Delta_m$$\geq$0.5~mag) $\sim$44$\%$. With the literature and our sample, we compute: f($\Delta_m$$\geq$0.7~mag) $\sim$22$\%$, f($\Delta_m$$\geq$0.6~mag) $\sim$39$\%$, f($\Delta_m$$\geq$0.5~mag)$\sim$42$\%$. It is important to mention that we are considering that the sparse lightcurves with low to moderate amplitude reported here are due to single objects. In some cases our temporal coverage is limited and some objects may be contact binaries with very slow rotation as 2014~JL$_{80}$. Therefore, our estimates are lower limits.    \\

Ten objects in our sample have absolute magnitude H$\geq$7.0~mag (diameter of 265/118~mag with albedo of 0.04/0.2). So far, observers have been focusing on large and bright TNOs, with only a handful of studies of smaller objects (see \citet{Thirouin2013} for a complete review). Therefore, we are testing a new size range of objects, and only little information is known about rotational properties of medium to small TNOs. In Figure~\ref{fig:all}, one can appreciate that the lightcurve amplitude is increasing at higher absolute magnitude (i.e. small objects). Therefore, we decide to test the percentage of high amplitude objects for smaller size. We follow the same approach as previously, but we now only focus on objects with H$\geq$6.0~mag (diameter of 419/188~mag assuming an albedo of 0.04/0.2). Based on our sample and the literature, we have: f($\Delta_m$$\geq$0.7~mag)$\sim$35$\%$, f($\Delta_m$$\geq$0.6~mag)$\sim$52$\%$, and f($\Delta_m$$\geq$0.5~mag)$\sim$50$\%$ using Equation~\ref{eq:frac1}. With Equation~\ref{eq:frac2}, we have: f($\Delta_m$$\geq$0.7~mag)$\sim$27$\%$, f($\Delta_m$$\geq$0.6~mag)$\sim$48$\%$, and f($\Delta_m$$\geq$0.5~mag)$\sim$52$\%$. In conclusion, $\sim$40$\%$ of the Plutinos and $\sim$50$\%$ of the small Plutinos could be contact binaries using a cut-off at $\Delta_m$$\geq$0.5~mag. 

Several studies mentioned that there is an anti-correlation between size and lightcurve amplitude \citep{Thirouin2010, Thirouin2013, Benecchi2013}. In other words, at smaller sizes (i.e. larger absolute magnitude), the lightcurve amplitude is larger suggesting that the small objects are more elongated and have more irregular shape. Such a tendency have been reported in all the dynamical group of TNOs \citep{Thirouin2013}. Therefore, one may expect that objects reported in this work are likely more elongated/deformed, however, their lightcurves are showing the U-/V-shapes indicatives of contact binary, and therefore the contact binary option is more likely.    

\subsection{Correlation/anti-correlation search}

We also search for correlation/anti-correlation between the lightcurve amplitude and the rotational period with the orbital elements (technique detailed in \citet{Thirouin2016}). Only two strong trends have been identified (Figure~\ref{fig:all}): the lightcurve amplitude is correlated with the absolute magnitude ($\rho$=0.544, significance level=98$\%$), and an anti-correlation between lightcurve amplitude and inclination ($\rho$=-0.505, significance level=97$\%$). The correlation between amplitude and size suggesting that the smaller objects present a large variability has been already noticed in several works and seems to be present in all the TNO dynamical groups \citep{Sheppard2008, Benecchi2013, Thirouin2013}. The anti-correlation between amplitude and inclination suggests that the variable objects are at low inclination. Such a trend has been noticed in several sub-populations of TNOs \citep{Thirouin2013}. However, it is interesting that the large amplitude objects are located at low inclination in the Plutino population. \citet{Sheppard2012} concluded that the 3:2 population presents a large variety of colors from neutral to ultra-red, suggesting a Cold Classical component at low inclination. Unfortunately, we do not have color studies for our objects to identify them as ultra-red objects (i.e. as a cold classical TNO) or not. However, color studies are available for 2001~QG$_{298}$ and 2000~GN$_{171}$ \citep{Sheppard2002, Sheppard2004}. With a B-R of 1.6~mag for 2001~QG$_{298}$ and 1.55~mag for 2000~GN$_{171}$, both objects are very red. Typical error bars are $\pm$0.04~mag, and thus taking into account this uncertainty, both objects are potentially ultra-red. In conclusion, both objects are potentially presenting similar characteristics as the dynamical Cold Classical TNOs. By studying the contact binaries in the Plutino and the Cold Classical populations, one will be able to check if both populations share the same binary fraction and characteristics, if there was some leakage of objects between the two populations, and constrain Neptune's migration (paper in preparation).


\section{Conclusions} 
 
We have displayed data of twelve Plutinos over two years using the Lowell's Discovery Channel Telescope. An homogeneous dataset reduced and analyzed the same way is presented. Our findings can be summarized as: 

\begin{itemize}

\item For nine objects, we report partial lightcurves showing a typical variability lower than 0.2~mag. Three objects in our sample of 12 are showing a lightcurve amplitude larger than 0.5~mag (25$\%$ of our sample). 2014~JL$_{80}$ is the slowest rotator in our sample with a period of 34.87~h and an amplitude of 0.55~mag. The period of 2014~JO$_{80}$ is 6.32~h and the amplitude is 0.60~mag. In the case of 2014~JQ$_{80}$, the lightcurve amplitude is 0.76~mag and the rotational period is 12.16~h. Lightcurves of 2014~JL$_{80}$, and 2014~JQ$_{80}$ display a U- and V-shape indicating contact binaries. In the case of the very slow rotator 2014~JL$_{80}$, the V-shape seems to be present, but unfortunately, we do not have the U-shape of the curve in our data.   

\item The large amplitude lightcurves are best explained by contact binary systems, but single very elongated objects cannot be ruled out. Assuming a contact binary configuration, we derive q=0.5, and a density of 1~g~cm$^{-3}$ for 2014~JL$_{80}$, a q=0.3 and a density of 3.25~g~cm$^{-3}$ for 2014~JO$_{80}$, and q=0.6 with a density of 1~g~cm$^{-3}$ for 2014~JQ$_{80}$.  

\item Thanks to this study, the current population of likely/confirmed contact binaries is composed of 8 TNOs: 1 dynamically Cold Classical object \citep{Thirouin2017b}, 1 in the Haumea family \citep{Lacerda2014}, 1 in the 5:2 mean motion resonance \citep{Thirouin2017} and 5 in the Plutino population (this work, and \citet{Sheppard2002, Sheppard2004}).    

\item Based on our sample, we estimate the fraction of contact binaries in the Plutino population. Using a cut-off of 0.5~mag, we find that $\sim$40$\%$ of the Plutinos can be contact binaries. Interestingly, all the known contact binaries in the Plutino population are small with absolute magnitude H$\geq$6~mag. We estimate a contact binary fraction of about 50$\%$ for the small Plutinos. This suggests the Plutino population may have significantly more contact binaries than other TNO populations. Future observations of smaller objects in other TNO populations are needed to confirm (or not) this finding.

\end{itemize}

%
 
\acknowledgments

Authors thank an anonymous referee for helpful comments and corrections. This research is based on data obtained at the Lowell Observatory's Discovery Channel Telescope (DCT). Lowell operates the DCT in partnership with Boston University, Northern Arizona University, the University of Maryland, and the University of Toledo. Partial support of the DCT was provided by Discovery Communications. LMI was built by Lowell Observatory using funds from the National Science Foundation (AST-1005313). We acknowledge the DCT operators: Andrew Hayslip, Heidi Larson, Teznie Pugh, and Jason Sanborn. A. Thirouin is partly supported by Lowell Observatory. Authors acknowledge support from the National Science Foundation, and the grant awarded to the ``Comprehensive Study of the Most Pristine Objects Known in the Outer Solar System" (AST-1734484).

 \clearpage

\onecolumn

\begin{table}[h!]
\caption{\label{Tab:Log_Obs} UT-Dates, heliocentric and geocentric (r$_{h}$, $\Delta$ respectively) distances and phase angle ($\alpha$, in degrees) are summarized.   }
\center
\begin{tabular}{lccccccc} 

\hline
Object & UT-date & Nb.   & r$_h$ &  $\Delta$ & $\alpha$   & Filter & Telescope\\
       &   &  of images   &  [AU]  &  [AU]  &  [$^{\circ}$]   & & \\
\hline
\hline
1995~HM$_{5}$&   &     &     &   &       &   &\\
& 05/20/2017  &  9   &  29.656   &  28.647 &  0.2 &  VR & DCT \\
\hline
(469362) 2001~KB$_{77}$&   &     &     &   &       &   &\\
& 06/18/2017  &   6  & 28.731 &  27.750 & 0.5  &  VR & DCT \\
\hline
2006~UZ$_{184}$&   &     &     &   &       &   &\\
& 11/30/2015  &  6   &   31.029  & 30.076-30.077  & 0.5  &  VR & DCT \\
& 12/01/2015  &  4   &  31.030 &  30.082 & 0.5 &  VR & DCT \\
\hline
2014~JL$_{80}$&   &     &     &   &       &   &\\
& 05/28/2017  &  10 &   28.563 & 27.578 &  0.5 &  VR & DCT \\
& 05/29/2017  & 18  & 28.563&   27.581-27.582  & 0.5 & VR & DCT \\
& 06/18/2017  & 7  & 28.559 &  27.706-27.707 & 1.1  &  VR & DCT \\
& 06/28/2017  &  6 &  28.557   & 27.806-27.807  &  1.4     &  VR & DCT \\
\hline
2014~JK$_{80}$&   &     &     &   &       &   &\\
& 06/27/2017  &  4   & 35.236    &  34.446-34.447 &  1.1 &  VR & DCT \\
& 06/28/2017  &  6   &  35.235   & 34.454-34.455  &  1.1     &  VR & DCT \\
\hline
2014~JO$_{80}$&   &     &     &   &       &   &\\
& 06/18/2017  &   13  &  31.938-31.939   &  31.008-31.009 &    0.7   &  VR & DCT \\
& 06/27/2017  &   4  &  31.945   & 31.078-31.079  &  1.0 &  VR & DCT \\
& 06/28/2017  &  12 & 31.946  & 31.086-31.087  &  1.0 &  VR & DCT \\
& 07/02/2017  &  13   &  31.949   &  31.124-31.126 &   1.1    &  VR & DCT \\
\hline
2014~JP$_{80}$&   &     &     &   &       &   &\\
& 06/27/2017  & 5    &  42.207   & 41.349  &  0.7 &  VR & DCT \\
\hline
2014~JQ$_{80}$&   &     &     &   &       &   &\\
& 05/01/2017  &   23  & 31.627 & 30.670-30.669 & 0.6  & VR  & DCT \\
& 05/20/2017  &  17   &  31.633   & 30.627-30.628  &   0.2    &  VR & DCT \\
& 05/28/2017  & 20  &   31.635  & 30.641-30.642 &  0.4 & VR  & DCT \\
& 05/29/2017  & 10 &  31.636 &  30.644-30.645 & 0.4 & VR  & DCT \\
& 06/18/2017  & 7 &  31.642 &  30.762-30.764 & 0.9  & VR  & DCT \\
& 06/27/2017  &6 & 31.645 & 30.850-30.851  & 1.2  & VR  & DCT \\
\hline
2014~JT$_{80}$&   &     &     &   &       &   &\\
& 06/27/2017  &  5   &  32.245   & 31.302-31.303  & 0.7 &  VR & DCT \\
\hline
2014~KC$_{102}$&   &     &     &   &       &   &\\
& 06/18/2017  &   4  &  30.505   & 29.526  &  0.5     &  VR & DCT \\
& 07/02/2017  &   15  & 30.498    & 29.581-29.582  & 0.8 & VR  & DCT \\
\hline
2014~KX$_{101}$&   &     &     &   &       &   &\\
& 05/28/2017  & 12  &  31.153   & 30.143 &  0.1 & VR  & DCT \\
& 05/29/2017  &  5   &  31.154   & 30.142  &  0.1  & VR  & DCT \\
\hline
2015~BA$_{519}$&   &     &     &   &       &   &\\
& 05/01/2017  &  16   & 31.041 &  30.053 &  0.4  &  VR & DCT \\
\hline
\hline

\end{tabular}
\end{table}
 
 \clearpage

\begin{scriptsize}
\begin{table}
\caption{\label{Summary_photo} Summary of this work. We report the preferred rotational period (P in h), the full lightcurve amplitude ($\Delta$m in mag.), and the Julian Date ($\varphi_{0}$, no light time correction) corresponding to the zero phase. Absolute magnitude (H), and the diameter considering an albedo of 0.04/0.2 are also indicated.     
}

\begin{tabular}{@{}lccc|cc|c} 
\hline
 Object  &  P & $\Delta$m &  $\varphi_{0}$ [JD] & H & Diameter & Contact binary?  \\
 
  &  [h]& [mag] & [2450000+] & [mag]  & [km] & \\
\hline
\hline
1995~HM$_{5}$ &  $>$0.6  &  $>$0.1  & 7893.81062 & 7.7 & 192/86 & \\
(469362) 2001~KB$_{77}$ &$>$4 &  $>$0.15 & 7922.69978& 7.4  & 220/98 &\\
2006~UZ$_{184}$ & $>$2 &  $>$0.2  & 7356.78287 &7.4  & 220/98 & \\
2014~JK$_{80}$ &  $>$1  &  $>$0.17  & 7931.71324  & 6.1& 400/179 &\\
2014~JL$_{80}$ &  34.87  &  0.55$\pm$0.03  & 7901.76240 & 7.1 & 253/113& Likely \\
2014~JO$_{80}$ & 6.32   &  0.60$\pm$0.05  & 7922.51572  & 7.4 & 220/98 &   Likely  \\
2014~JP$_{80}$ & $>$0.5 & $>$0.1  & 7931.85876 & 4.9 & 696/311 & \\
2014~JQ$_{80}$ &  12.16  &  0.76$\pm$0.04  & 7874.74484  &7.0 &  265/118 & Likely\\
2014~JT$_{80}$ & $>$0.8 & $>$0.1 &7931.76831    &7.1 &253/113 & \\
2014~KC$_{102}$ & $>$4.5   & $>$0.2  & 7922.70464 &7.1 & 253/113& \\
2014~KX$_{101}$ &  $>$3  &  $>$0.2  & 7901.78398 &7.4 & 220/98 & \\
2015~BA$_{519}$ &  $>$4/8$^{a}$  &  $\sim$0.16  & 7874.73116 & 7.4 & 220/98 & \\
\hline\hline
\end{tabular}
\\
Notes:\\
In case of partial lightcurve, the lower limit to the period reported is the duration of our observing block. \\
$^{a}$: The lightcurve of 2015~BA$_{519}$ shows a maximum and minimum over about 4~h of observations. Therefore, the rotational period of this object is likely $>$8~h (double-peaked).
\end{table}
\end{scriptsize}

\clearpage

\begin{deluxetable}{lcccccc}
\tabletypesize{\scriptsize}
\tablewidth{0pt}
\tablecaption{\label{Tab:photoplutino} Short-term variability of the Plutino trans-Neptunian objects is summarized. Preferred rotational period is indicated in bold. Reference list is after the table.    }
\tablehead{
\colhead{Object}           & \colhead{Single-peaked period [h]}   
 & \colhead{Double-peaked period [h]}  &
\colhead{$\Delta m$ [mag]}          & \colhead{H [mag]}    &
\colhead{Ref.}}
\startdata
(134340) Pluto &   153.2 & - & 0.33 & -0.7 & B97\\
Charon &   153.6 & - & 0.08 & 0.9 & B97\\
(15789) 1993~SC  & 7.7129 & - & 0.5 & 7.0 & W95\\
& - & - & $<$0.2 & ... & T97, D97$^{a}$\\
(15810) 1994~JR$_{1}$ Arawn & - &  5.47$\pm$0.33  & 0.58 & 7.7 & P16 \\
(15820) 1994~TB &   3.0/3.5 & 6.0/7.0 & 0.26/0.34 & 7.3 & RT99\\
&  - & - & $<$0.04 & ... & SJ02\\
(32929) 1995~QY$_{9}$&   -  & 7.3$\pm$0.1 & 0.60$\pm$0.04 & 8.0 & RT99,SJ02\\
(15875) 1996~TP$_{66}$&  1.96 & - & $<$0.04 & 7.0 & CB99$^b$\\
&   - & - & $<$0.12 & ... & RT99\\
(118228) 1996~TQ$_{66}$&   - & - & $<$0.22 & 6.9 & RT99\\
(91133) 1998~HK$_{151}$&   - & - & $<$0.15 & 7.6 & SJ02\\
(33340) 1998~VG$_{44}$ &   - & - & $<$0.10 & 6.5 & SJ02\\
(47171) 1999~TC$_{36}$ & 6.21$\pm$0.02 & - & 0.06 & 4.9 & O03\\
&   - & - & $<$0.07 & ... & LL06\\
&   - & - & $<$0.05 & ... & SJ03\\
(38628) 2000~EB$_{173}$ Huya& (6.68/\textbf{6.75}/6.82)$\pm$0.01 & - & $<$0.1 & 4.8 & O03\\
&   - & - & $<$0.15 & ... & SJ02\\
&   - & - & $<$0.06 & ... & SR02\\
&   - & - & $<$0.04 & ... & SJ03,LL06\\
&   5.21 & - & 0.02$\pm$0.01 & ... & T14\\
&   4.45$\pm$0.07 & - & $\sim$0.1 & ... & G16\\
2000~FV$_{53}$ &   - & 7.5 & 0.07$\pm$0.02 & 8.3 & TB06\\
(47932) 2000~GN$_{171}$ & - & 8.329$\pm$0.005 & 0.61$\pm$0.03 & 6.2 & SJ02\\
         &   - & - &  0.53 &  ... & D08$^{c}$\\
         &  - & - & 0.64$\pm$0.11 & ... & R07\\ 
2001~KD$_{77}$&  - & - & $<$0.07 & 5.8 & SJ03\\
(469372) 2001~QF$_{298}$&   - & - & $<$0.12 & 5.2 & SJ03\\
&   - & - & $\sim$0.11 & ... & T13 \\
(139775) 2001~QG$_{298}$&   6.8872$\pm$0.0002 & 13.7744$\pm$0.0004 & 1.14$\pm$0.04 & 6.9 & SJ04\\
  & - & 13.7744$\pm$0.0004 & 0.7$\pm$0.01 & ... & L11\\
(28978) 2001~KX$_{76}$ Ixion &   - & - & $<$0.05 & 3.6 & O03,SJ03\\
&  15.9$\pm$0.5 & - & 0.06$\pm$0.03 & ... & RP10\\
&  12.4$\pm$0.3 & - & - & ... & G16\\
(55638) 2002~VE$_{95}$ &   (6.76/6.88/7.36/9.47)$\pm$0.01 & - & 0.08$\pm$0.04 & 5.5 & O06\\
&  - & - & $<$0.06 & ... & SJ03\\
&   9.97 & - & 0.05$\pm$0.01 & ... & T10\\
(208996) 2003~AZ$_{84}$&  (4.32/5.28/6.72/\textbf{6.76})$\pm$0.01 & - & 0.10$\pm$0.04 & 3.7 & O06\\
&   6.72$\pm$0.05 & - & 0.14$\pm$0.03 & ... & SJ03\\
&   6.79 & - & 0.07$\pm$0.01 & ... & T10\\
&   6.78 & - & 0.07$\pm$0.01 & ... & T14\\
2003~HA$_{57}$ &   -  & 6.44 & 0.31$\pm$0.03   & 8.1 &  T16 \\ 
(455502) 2003~UZ$_{413}$ &  - & 4.13$\pm$0.05 & 0.13$\pm$0.03 & 4.3 & P09\\
(84922) 2003~VS$_{2}$ & (3.71 or 4.39)$\pm$0.01 & - & 0.23$\pm$0.07 & 4.2 & O06\\
&   - & 7.41$\pm$0.02 & 0.21$\pm$0.02 & ... & S07\\
&   - & 7.42 & 0.21$\pm$0.01 & ... & T10\\
&   - & 7.4208 & 0.224$\pm$0.013 & ... & T13\\
(90482) 2004~DW Orcus &  7.09/\textbf{10.08$\pm$0.01}/17.43 & \textbf{20.16} & 0.04$\pm$0.02 & 2.2 & O06\\
&  13.19 & - & 0.18$\pm$0.08 & ... & R07\\
&  - & - & $<$0.03 & ... & S07\\
&  10.47 & - & 0.04$\pm$0.01 & ... & T10\\
&  11.9$\pm$0.5 & - & - & ... & G16\\
(469708) 2005~GE$_{187}$ & 6.1 & - & 0.5 & 7.3 & S10\\ 
 	     &  - &  11.99 &   0.29$\pm$0.02  &  ... &  T16\\
(469987) 2006~HJ$_{123}$ &  - & - & $<$0.13 & 5.9 & BS13\\
(341520) 2007~TY$_{430}$ Mors-Somnus& - & 9.28 & 0.24$\pm$0.05 & 6.9 & T14$^{d}$ \\
\enddata
\\
Notes: \\
$^a$: Results from \citet{Williams1995} were not supported by \citet{Tegler1997}, and \citet{Davies1997}. Thus, this object will not be considered in our study. 
$^b$: Rotational period of 1996~TP$_{66}$ is likely wrong, and will not be used in this work. 
$^c$: \citet{Dotto2008} used the rotational period from \citet{Sheppard2002} to fit their data. They reported a lightcurve amplitude of 0.60$\pm$0.03~mag, but based on their fit (Figure~3 of \citet{Dotto2008}), the lightcurve amplitude is about 0.53~mag.
$^{d}$: According to \citet{Sheppard2012b}, 2007~TY$_{430}$ is dynamically cold classical object in the Plutino population.  \\
References list: \\
W95: \citet{Williams1995}; B97: \citet{Buie1997}; D97:\citet{Davies1997}, T97: \citet{Tegler1997}; CB99: \citet{Collander1999}; RT99: \citet{Romanishin1999}; SJ02: \citet{Sheppard2002}; SR02: \citet{Schaefer2002}; O03: \citet{Ortiz2003}; SJ03: \citet{Sheppard2003}; SJ04: \citet{Sheppard2004}; LL06: \citet{Lacerda2006}; O06: \citet{Ortiz2006}; TB06: \citet{Trilling2006}; R07: \citet{Rabinowitz2007}; S07: \citet{Sheppard2007}; D08: \citet{Dotto2008}; P09: \citet{Perna2009}; RP10: \citet{Rousselot2010}; S10: \citet{Snodgrass2010}; T10: \citet{Thirouin2010}; L11: \citet{Lacerda2011}; BS13: \citet{Benecchi2013}; T13: \citet{Thirouin2013}; T14: \citet{Thirouin2014}; G16: \citet{Galiazzo2016}; P16: \citet{Porter2016}; T16: \citet{Thirouin2016}.
 
\end{deluxetable}

 \clearpage

 \clearpage

 \clearpage

\clearpage

\clearpage

\begin{figure*}
\includegraphics[width=9cm, angle=0]{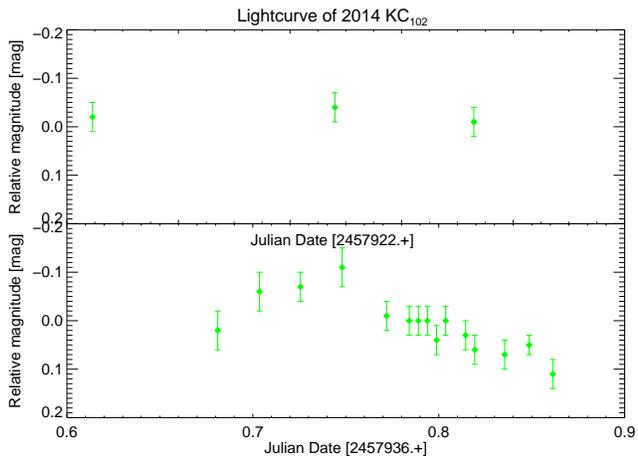}
\includegraphics[width=9cm, angle=0]{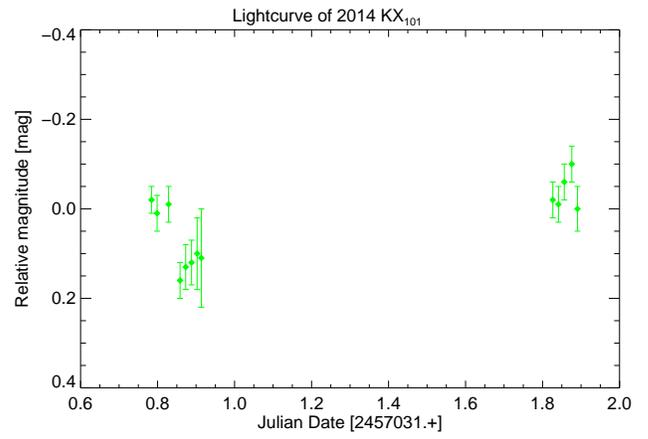}
\includegraphics[width=9cm, angle=0]{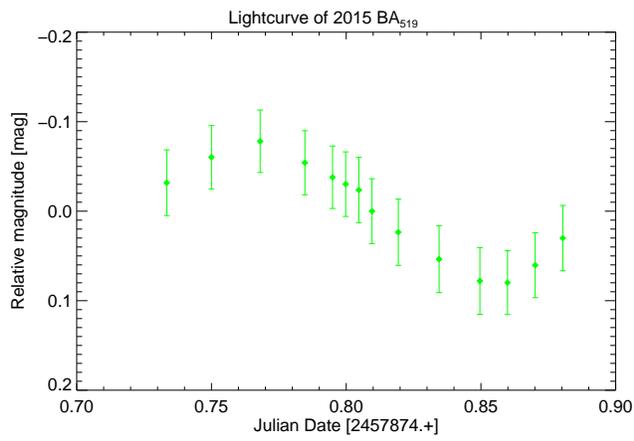}
\caption{\textit{Partial lightcurves of several Plutinos:} continuation of Figure~5. }
\label{fig:LCnocb2}
\end{figure*}

\clearpage

\begin{figure*}
\includegraphics[width=12cm, angle=0]{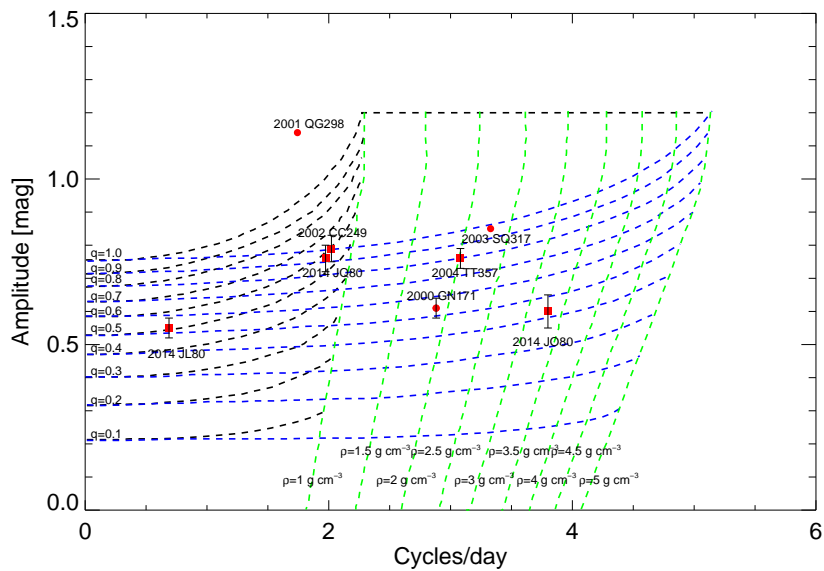}
\caption{\textit{The network of Roche sequences}: A complete description of this plot is available in \citet{Leone1984, Thirouin2017}. Confirmed/likely contact binaries from this work and the literature are plotted. Name of each objects is also indicated. }
\label{fig:Roche}
\end{figure*}

\begin{figure*}
\includegraphics[width=13.5cm, angle=180]{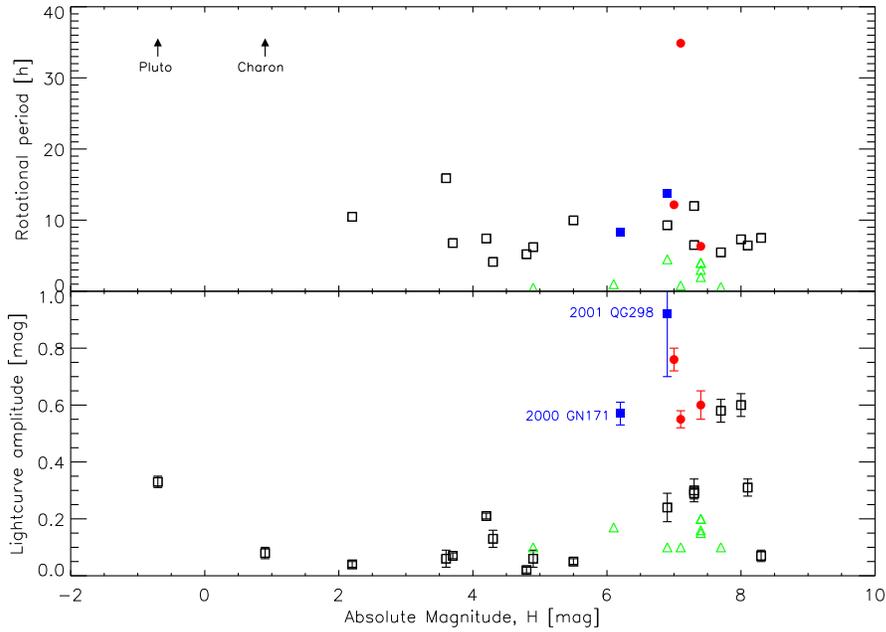}
\includegraphics[width=13.5cm, angle=180]{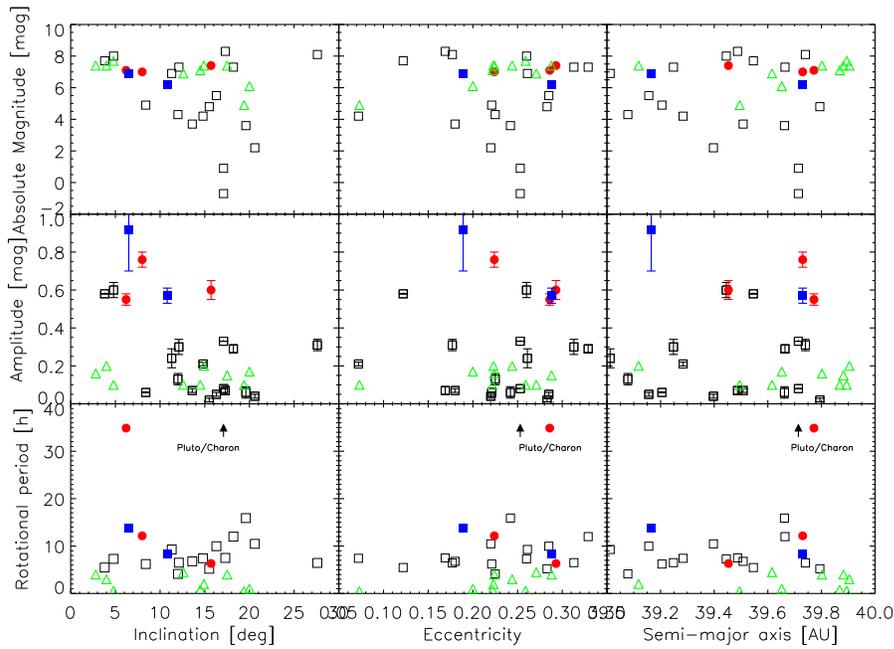}
\caption{\textit{Rotational properties and orbital elements}: All the Plutinos with published lightcurves are plotted. Open black squares are from the literature, green triangles are plutinos with partial lightcurves reported in this work (lower estimates for the period and amplitude), red circles are the likely contact binaries from this work and blue circles are the likely/confirmed contact binaries from the literature. There is a correlation between amplitude and absolute magnitude and an anti-correlation between inclination and lightcurve amplitude.       }
\label{fig:all}
\end{figure*}
 
 \clearpage

\appendix
\section*{Appendix A}
\setcounter{section}{1}
\begin{deluxetable}{lccc}
 \tablecaption{\label{Tab:Summary_photo} Photometry used in this paper is available in the following table.  }
\tablewidth{0pt}
\tablehead{
Object  & JD & Rel. mag. &  Err.  \\
        &          &   [mag]           &  [mag]\\
}
\startdata  
1995~HM$_{5}$ &   &   &   \\
 & 2457893.81062 & 0.03 & 0.04 \\ 
& 2457893.82056 & 0.00 & 0.04 \\ 
& 2457893.83067 & -0.08 & 0.01 \\ 
& 2457893.84067 & -0.05 & 0.02 \\ 
& 2457893.85073 & -0.04 & 0.03 \\
 & 2457893.86061 & -0.04 & 0.01 \\ 
& 2457893.87065 & 0.00 & 0.02 \\ 
& 2457893.88050 & 0.03 & 0.05 \\ 
& 2457893.89038 & 0.00 & 0.03 \\ 
\hline
2001~KB$_{77}$ &   &   &   \\
 & 2457922.69978 & 0.00 & 0.03 \\ 
& 2457922.74331 & 0.00 & 0.03 \\
 & 2457922.76815 & 0.07 & 0.04 \\ 
& 2457922.83814 & 0.03 & 0.03 \\ 
& 2457922.85305 & -0.06 & 0.03 \\ 
& 2457922.86298 & -0.08 & 0.03 \\
\hline
2006~UZ$_{184}$ &   &   &   \\
 & 2457356.78287 & -0.08 & 0.05 \\
& 2457356.79985 & -0.13 & 0.05 \\ 
& 2457356.81682 & -0.07 & 0.06 \\ 
& 2457356.83377 & -0.03 & 0.06 \\ 
& 2457356.85215 & 0.03 & 0.07 \\ 
& 2457356.86895 & -0.06 & 0.07 \\ 
& 2457357.83488 & 0.09 & 0.12 \\ 
& 2457357.85205 & 0.14 & 0.11 \\ 
& 2457357.86094 & 0.25 & 0.12 \\ 
& 2457357.87037 & 0.24 & 0.12 \\
\hline
2014~JL$_{80}$ &   &   &   \\
 & 2457901.76402 & 0.14 & 0.03 \\
 & 2457901.77899 & 0.15 & 0.03 \\ 
& 2457901.79376 & 0.15 & 0.03 \\
 & 2457901.80857 & 0.14 & 0.03 \\
 & 2457901.82354 & 0.07 & 0.03 \\ 
& 2457901.83837 & 0.03 & 0.03 \\ 
& 2457901.85321 & 0.01 & 0.03 \\ 
& 2457901.86809 & -0.01 & 0.03 \\
 & 2457901.88304 & -0.05 & 0.03 \\
 & 2457901.89786 & -0.02 & 0.05 \\ 
& 2457902.82155 & -0.28 & 0.03 \\
& 2457902.83643 & -0.29 & 0.03 \\
 & 2457902.85129 & -0.30 & 0.03 \\ 
& 2457902.86610 & -0.25 & 0.03 \\ 
& 2457902.87091 & -0.29 & 0.03 \\ 
& 2457902.88573 & -0.23 & 0.03 \\ 
& 2457902.90052 & -0.19 & 0.03 \\ 
& 2457902.91029 & -0.21 & 0.04 \\
 & 2457922.68985 & 0.10 & 0.03 \\
 & 2457922.73343 & 0.24 & 0.03 \\ 
& 2457922.75813 & 0.20 & 0.03 \\
 & 2457922.78302 & 0.19 & 0.03 \\ 
& 2457922.79789 & 0.17 & 0.03 \\ 
& 2457922.81280 & 0.10 & 0.03 \\ 
& 2457922.82776 & 0.05 & 0.03 \\
 & 2457932.70192 & -0.16 & 0.02 \\ 
& 2457932.71803 & -0.13 & 0.02 \\ 
& 2457932.73411 & -0.11 & 0.02 \\
 & 2457932.75037 & -0.08 & 0.02 \\ 
& 2457932.77251 & 0.04 & 0.03 \\
\hline
2014~JK$_{80}$ &   &   &   \\
 & 2457931.71324 & 0.01 & 0.03 \\ 
& 2457931.73529 & 0.11 & 0.02 \\ 
& 2457931.75585 & 0.12 & 0.02 \\ 
& 2457931.86866 & -0.03 & 0.03 \\
 & 2457932.67154 & 0.08 & 0.03 \\ 
& 2457932.69694 & 0.04 & 0.02 \\ 
& 2457932.71301 & -0.01 & 0.02 \\ 
& 2457932.72915 & -0.03 & 0.02 \\ 
& 2457932.74536 & -0.02 & 0.02 \\
 & 2457932.76141 & -0.05 & 0.02 \\
\hline
2014~JO$_{80}$ &   &   &   \\
 & 2457922.51572 & -0.22 & 0.04 \\ & 2457922.55931 & 0.04 & 0.04 \\ & 2457922.58397 & 0.21 & 0.05 \\ & 2457922.60888 & 0.02 & 0.05 \\ & 2457922.62380 & -0.11 & 0.04 \\ & 2457922.63871 & -0.10 & 0.04 \\ & 2457922.65367 & -0.08 & 0.04 \\ & 2457922.66899 & -0.09 & 0.05 \\ & 2457922.67890 & -0.02 & 0.06 \\ & 2457922.68883 & 0.05 & 0.06 \\ & 2457922.69364 & -0.03 & 0.06 \\ & 2457922.69845 & 0.16 & 0.08 \\ & 2457922.70355 & 0.02 & 0.07 \\ & 2457931.59498 & -0.10 & 0.04 \\ & 2457931.61365 & -0.06 & 0.04 \\ & 2457931.62604 & 0.00 & 0.04 \\ & 2457931.67432 & 0.06 & 0.06 \\ & 2457932.48703 & -0.08 & 0.06 \\ & 2457932.51189 & -0.08 & 0.04 \\ & 2457932.52791 & -0.14 & 0.03 \\ & 2457932.54409 & -0.11 & 0.03 \\ & 2457932.56027 & 0.03 & 0.03 \\ & 2457932.57636 & 0.20 & 0.04 \\ & 2457932.58742 & 0.30 & 0.04 \\ & 2457932.59845 & 0.22 & 0.04 \\ & 2457932.60441 & 0.17 & 0.04 \\ & 2457932.61036 & 0.04 & 0.03 \\ & 2457932.61633 & -0.07 & 0.03 \\ & 2457932.62227 & -0.07 & 0.03 \\ & 2457936.48638 & -0.18 & 0.08 \\ & 2457936.49655 & 0.01 & 0.09 \\ & 2457936.50709 & 0.10 & 0.09 \\ & 2457936.51822 & 0.11 & 0.09 \\ & 2457936.52931 & 0.39 & 0.10 \\ & 2457936.54040 & 0.13 & 0.08 \\ & 2457936.55159 & 0.09 & 0.07 \\ & 2457936.56266 & 0.11 & 0.07 \\ & 2457936.57412 & -0.02 & 0.08 \\ & 2457936.58614 & -0.13 & 0.06 \\ & 2457936.59847 & -0.18 & 0.06 \\ & 2457936.62931 & -0.23 & 0.05 \\ 
\hline
2014~JP$_{80}$ & & & \\
 & 2457931.85879 & 0.01 & 0.02 \\ 
& 2457931.87378 & 0.02 & 0.02 \\ 
& 2457931.87801 & 0.00 & 0.02 \\ 
& 2457931.88222 & -0.01 & 0.02 \\ 
& 2457931.88645 & -0.01 & 0.02 \\
\hline
2014~JQ$_{80}$ &   &   &   \\
 & 2457874.56771 & 0.28 & 0.05 \\ & 2457874.58431 & 0.26 & 0.04 \\ & 2457874.61256 & 0.01 & 0.03 \\ & 2457874.63733 & -0.06 & 0.03 \\ & 2457874.64747 & -0.12 & 0.03 \\ & 2457874.66268 & -0.13 & 0.03 \\ & 2457874.67782 & -0.12 & 0.02 \\ & 2457874.68806 & -0.14 & 0.02 \\ & 2457874.69826 & -0.09 & 0.03 \\ & 2457874.70837 & -0.05 & 0.03 \\ & 2457874.71855 & -0.01 & 0.03 \\ & 2457874.72859 & 0.01 & 0.03 \\ & 2457874.73764 & 0.10 & 0.03 \\ & 2457874.74648 & 0.21 & 0.03 \\ & 2457893.63097 & -0.06 & 0.03 \\ & 2457893.64101 & -0.04 & 0.03 \\ & 2457893.65111 & -0.10 & 0.04 \\ & 2457893.66115 & -0.13 & 0.03 \\ & 2457893.67116 & -0.19 & 0.03 \\ & 2457893.68113 & -0.21 & 0.03 \\ & 2457893.69109 & -0.25 & 0.03 \\ & 2457893.70102 & -0.17 & 0.03 \\ & 2457893.71090 & -0.17 & 0.03 \\ & 2457893.72083 & -0.08 & 0.04 \\ & 2457893.72562 & 0.00 & 0.04 \\ & 2457893.73051 & -0.01 & 0.04 \\ & 2457893.73529 & 0.00 & 0.04 \\ & 2457893.74010 & 0.05 & 0.05 \\ & 2457893.74489 & 0.08 & 0.04 \\ & 2457893.74970 & 0.07 & 0.05 \\ & 2457893.75449 & 0.13 & 0.05 \\ & 2457901.48354 & -0.02 & 0.03 \\ & 2457901.49326 & -0.04 & 0.03 \\ & 2457901.50170 & -0.09 & 0.03 \\ & 2457901.51366 & -0.08 & 0.03 \\ & 2457901.52225 & -0.12 & 0.03 \\ & 2457901.53065 & -0.19 & 0.03 \\ & 2457901.53909 & -0.10 & 0.03 \\ & 2457901.54762 & -0.11 & 0.03 \\ & 2457901.55603 & -0.11 & 0.03 \\ & 2457901.56440 & -0.08 & 0.03 \\ & 2457901.58213 & 0.01 & 0.03 \\ & 2457901.59716 & 0.07 & 0.03 \\ & 2457901.61194 & 0.27 & 0.03 \\ & 2457901.62676 & 0.35 & 0.03 \\ & 2457901.64163 & 0.53 & 0.04 \\ & 2457901.65651 & 0.51 & 0.04 \\ & 2457901.67137 & 0.47 & 0.04 \\ & 2457901.68621 & 0.23 & 0.03 \\ & 2457901.70116 & 0.09 & 0.04 \\ & 2457901.71594 & 0.03 & 0.04 \\ & 2457902.63970 & 0.33 & 0.04 \\ & 2457902.65451 & 0.45 & 0.05 \\ & 2457902.68421 & 0.51 & 0.06 \\ & 2457902.70386 & 0.28 & 0.05 \\ & 2457902.71862 & 0.21 & 0.05 \\ & 2457902.72843 & 0.02 & 0.05 \\ & 2457902.73822 & -0.06 & 0.05 \\ & 2457902.74311 & -0.03 & 0.05 \\ & 2457902.74794 & -0.03 & 0.05 \\ & 2457922.50729 & -0.17 & 0.03 \\ & 2457922.55085 & -0.12 & 0.03 \\ & 2457922.57560 & -0.09 & 0.03 \\ & 2457922.60046 & 0.12 & 0.04 \\ & 2457922.61527 & 0.13 & 0.04 \\ & 2457922.63021 & -0.06 & 0.03 \\ & 2457922.64512 & 0.28 & 0.04 \\ & 2457931.51794 & 0.33 & 0.11 \\ & 2457931.52869 & 0.74 & 0.12 \\ & 2457931.55075 & 0.63 & 0.06 \\ & 2457931.56522 & 0.53 & 0.04 \\ & 2457931.58388 & 0.44 & 0.04 \\ & 2457931.60254 & 0.32 & 0.04 \\ 
\hline
2014~JT$_{80}$ &   &   &   \\
 & 2457931.76831 & 0.00 & 0.03 \\ 
& 2457931.78695 & -0.10 & 0.03 \\ 
& 2457931.79933 & 0.03 & 0.03 \\
\hline
2014~KC$_{102}$ &   &   &   \\
 & 2457922.70464 & -0.02 & 0.03 \\ 
& 2457922.74820 & -0.04 & 0.03 \\ 
& 2457922.77305 & -0.01 & 0.03 \\ 
& 2457936.68128 & 0.02 & 0.04 \\
 & 2457936.70355 & -0.06 & 0.04 \\
 & 2457936.72572 & -0.07 & 0.03 \\
 & 2457936.74803 & -0.11 & 0.04 \\ 
& 2457936.77202 & -0.01 & 0.03 \\ 
& 2457936.78435 & 0.00 & 0.03 \\
 & 2457936.78916 & 0.00 & 0.03 \\ 
& 2457936.79394 & 0.00 & 0.03 \\
 & 2457936.79875 & 0.04 & 0.03 \\ 
& 2457936.80354 & 0.00 & 0.03 \\ 
& 2457936.81460 & 0.03 & 0.03 \\ 
& 2457936.81941 & 0.06 & 0.03 \\ 
& 2457936.83553 & 0.07 & 0.03 \\ 
& 2457936.84842 & 0.05 & 0.02 \\ 
& 2457936.86129 & 0.11 & 0.03 \\
\hline
2014~KX$_{101}$ &   &   &   \\
 & 2457901.78398 & -0.02 & 0.03 \\
 & 2457901.79876 & 0.01 & 0.04 \\ 
& 2457901.82852 & -0.01 & 0.04 \\ 
& 2457901.85819 & 0.16 & 0.04 \\
 & 2457901.87313 & 0.13 & 0.05 \\ 
& 2457901.88802 & 0.12 & 0.05 \\ 
& 2457901.90293 & 0.10 & 0.08 \\ 
& 2457901.91350 & 0.11 & 0.11 \\ 
& 2457902.82649 & -0.02 & 0.04 \\
 & 2457902.84142 & -0.01 & 0.04 \\
 & 2457902.85625 & -0.06 & 0.04 \\
 & 2457902.87591 & -0.10 & 0.04 \\
 & 2457902.89068 & 0.00 & 0.05 \\
\hline
2015~BA$_{519}$ &   &   &   \\    
 &2457874.73348 & -0.03 & 0.04 \\
 & 2457874.74995 & -0.06 & 0.04 \\ 
& 2457874.76792 & -0.08 & 0.03 \\ 
& 2457874.78450 & -0.05 & 0.04 \\ 
& 2457874.79498 & -0.04 & 0.03 \\
 & 2457874.79977 & -0.03 & 0.04 \\ 
& 2457874.80458 & -0.02 & 0.04 \\ 
& 2457874.80936 & 0.00 & 0.04 \\ 
& 2457874.81954 & 0.02 & 0.04 \\
 & 2457874.83469 & 0.05 & 0.04 \\ 
& 2457874.84984 & 0.08 & 0.04 \\
 & 2457874.86009 & 0.08 & 0.04 \\ 
& 2457874.87030 & 0.06 & 0.04 \\
 & 2457874.88044 & 0.03 & 0.04 \\
 & 2457874.90070 & 0.00 & 0.04 \\
\enddata

\end{deluxetable}

\end{document}